\documentclass[11pt,prd,notitlepage,tightenlines,nofootinbib,superscriptaddress]{revtex4-1}
\usepackage[utf8]{inputenc}
\usepackage{rotating}
\usepackage{amsfonts}
\usepackage{amsthm}
\usepackage{amssymb,hyperref,amsmath}
\usepackage{mathtools}
\usepackage{color}
\usepackage{graphicx}
\usepackage{multirow}
\usepackage{float}
\usepackage[caption=false]{subfig}
\usepackage{soul}
\usepackage{hepunits}
\newcommand*\diff{\mathop{}\!\mathrm{d}}

\usepackage{braket}
\usepackage{bm}
\usepackage{txfonts}
\usepackage{mathrsfs} 
\begin{document}
 \bibliographystyle{apsrev4-1}
 \title{Radiative transitions between $0^{-+}$ and $1^{--}$ heavy quarkonia on the light front}
\author{Meijian Li}
\email{meijianl@iastate.edu}
\affiliation{Department of Physics and Astronomy, Iowa State University, Ames, IA, 50014}
\author{Yang Li}
\email{Corresponding author: leeyoung@iastate.edu}
\affiliation{Department of Physics, College of William \& Mary, Williamsburg, VA, 23185}
\author{Pieter Maris}
\email{pmaris@iastate.edu}
\affiliation{Department of Physics and Astronomy, Iowa State University, Ames, IA, 50014}
\author{James P. Vary}
\email{jvary@iastate.edu}
\affiliation{Department of Physics and Astronomy, Iowa State University, Ames, IA, 50014}
\date{\today}
\begin{abstract}
  We present calculations of radiative transitions between vector and pseudoscalar quarkonia in the light-front Hamiltonian
  approach. The valence sector light-front wavefunctions of heavy quarkonia are obtained from the Basis Light-Front Quantization (BLFQ) approach
  in a holographic basis. We study the transition form factor with both the traditional ``good current'' $J^+$ and the transverse
  current $\vec J_\perp$
 (in particular, $J^R=J^x+i J^y$). This allows us to investigate the role of rotational symmetry by considering vector mesons with different magnetic
  projections ($m_j=0,\pm 1$). We use the $m_j=0$ state of the vector meson to obtain the transition form factor,
  since this procedure employs the dominant spin
  components of the light-front wavefunctions and is more robust in practical calculations. While the $m_j=\pm 1$ states are also
  examined, transition form factors depend on subdominant components of the light-front wavefunctions and are less robust. Transitions between states below the open-flavor thresholds are computed, including those for excited states. Comparisons are made with the experimental measurements as well as with Lattice QCD and quark model results.
 In addition, we apply the transverse current to calculate the decay constant of vector mesons where we obtain consistent results using either $m_j=0$ or
 $m_j=1$ light-front wavefunctions. This consistency provides evidence for features of rotational symmetry within the model.
\end{abstract}

\maketitle
\section{Introduction}

Radiative transitions offer insights into the internal structure of quark-antiquark bound states through electromagnetic
probes. The radiative transition between $0^{-+}$  (pseudoscalar) and $1^{--}$ (vector) mesons via the emission of a photon is characterized as the magnetic dipole (M1) transition.
This transition mode is known to be sensitive to relativistic
effects~\cite{PhysRevD.64.074011}, especially for those between different spatial multiplets, such as $\eta_c(2S)\to
J/\psi(1S)+\gamma$.

Heavy quarkonium is often dubbed as the ``hydrogen atom'' of quantum chromodynamics (QCD). It provides an ideal testing ground for
various investigations to understand QCD. States below the open flavor threshold ($D\overline{D}$ for charmonium and $B\overline B$ for
bottomonium) have very narrow widths since they cannot decay via any Okubo-Zweig-Iizuka allowed strong
decay channels~\cite{OZI1,*OZI2,*OZI3}. Electromagnetic transition rates are therefore important. Comparing the theoretical and
experimental rates for radiative transitions then provides guidance to improve our understanding of the internal
structure of heavy quarkonia.
Recently, some of us proposed a model based on the light-front holographic QCD and the one-gluon exchange with a running
coupling~\cite{Yang_fix, Yang_run, Shuo_Bc}.
In this model, charmonia and bottomonia are solved as relativistic quark-antiquark
bound states using Basis Light-Front Quantization (BLFQ), a nonperturbative Hamiltonian approach within light-front
dynamics~\cite{1stBLFQ}.
The model provides a reasonable description of the mass spectrum and other properties that were studied.
Observables including decay constants, r.m.s. radii~\cite{YL_radii}, distribution amplitudes and parton distributions as well as diffractive vector meson productions~\cite{Chen:VM} have been
directly calculated from the
light-front wavefunctions (LFWFs), and are in reasonable agreement with experiments and with other approaches (see also
Ref.~\cite{Lekha_EFF}).
Therefore, we are motivated to investigate radiative transitions within this model. On the one hand, we
hope to further test the model by comparing with the existing experimental results. On the other hand, results calculated for
transitions that have not yet been measured provide predictions for future experiments.

In this work, we derive the formulae for radiative transitions between $0^{-+}$ and $1^{--}$ mesons on the light
front, using both the traditional ``good current'' $J^+=J^0+J^z$ and the transverse current $\vec J_\perp=(J^x, J^y)$. Though, in
principle, these two choices should be equivalent due to Lorentz covariance, adoption of certain approximations
in the model may lead to violation of the Lorentz symmetry that would be evident through inequivalent results.
In nonrelativistic quantum mechanics, magnetic moments and transitions can only be extracted from the \emph{current density} $\vec
J =(J^x, J^y, J^z)$ rather than the \emph{charge density} $J^0$. Therefore one may expect that for the M1 transitions in nonrelativistic
systems such as heavy quarkonia, the transverse current density  $\vec J_\perp$
could be better than the charge density $J^+$.
Specifically, as we will see later, the transverse current allows us to extract the transition form factor through the $m_j=0$ state of the vector
meson, which is not accessible with $J^+$. The calculation with the $m_j=0$ state provides a more robust result by employing the
dominant spin components of the two mesons (spin-triplet for the vector and
spin-singlet for the pseudoscalar) in the transition, while that with $m_j=\pm 1$ always requires subdominant components of the
LFWFs with relativistic origins.
So in this work, we obtain the transition form factors with the $m_j=0$ states of the vector mesons
through the transverse current. The results from $m_j=1$, though less robust, are also presented for comparison.
 In addition, as a cross check, we revisit the decay constants by utilizing the transverse current to compare with the previous calculation using $J^+$ in Ref.~\cite{Yang_run}. It follows that different $m_j$ components of the vector mesons are involved. This provides us with a different yet pertinent perspective to understand the degree of  Lorentz symmetry manifestation in the current model.

The layout of the paper is as follows. In Sec.~\ref{sec:TFF}, we introduce the formalism and methods to calculate the M1 transition
form factor on the light front. In Sec.~\ref{sec:BLFQ}, we apply the formalism to heavy quarkonia in the BLFQ approach.
Sec.~\ref{sec:result} presents the numerical results of transition form factors. In Sec.~\ref{sec:cons}, we perform the calculations of vector meson decay constants with different magnetic projections. We conclude the paper in
Sec.~\ref{sec:summary}.

\section{Transition form factors on the light front}\label{sec:TFF}
\subsection{Transition form factor and decay width}
The matrix element for the radiative transition between a vector meson ($\mathcal{V}$) with four-momentum $P$ and polarization $m_j$
and a pseudoscalar  ($\mathcal{P}$) with four-momentum $P'$ via emission of a photon,
can be parametrized in terms of the transition form factor $V(Q^2)$ as~\cite{Dudek_JPsi},
\begin{align}
  \label{eq:Vq2_def}
   I^\mu_{m_j}\equiv
  \bra{\mathcal{P}(P')} J^\mu(0)\ket{\mathcal{V} (P,m_j)}
  =\frac{2 V(Q^2)}{m_{\mathcal{P}}+m_{\mathcal{V}}}\epsilon^{\mu\alpha\beta\sigma} {P'}_\alpha P_\beta  e_{\sigma}(P, m_j)
  \;,
\end{align}
where we define $Q^2\equiv -q^2$, with $q^\mu = {P'}^\mu - P^\mu$ representing the four-momentum of the photon. $m_{\mathcal{P}}$
and $m_{\mathcal{V}}$ are the masses of the pseudoscalar and the vector, respectively.
$e_{\sigma}$ is the polarization vector of the vector meson. $J^\mu(x)$ is the current operator.

In the physical process of $\mathcal{V} \to \mathcal{P}+\gamma$, the photon is on shell ($Q^2=0$). The transition amplitude is:
\begin{align}\label{eq:amplitude}
  \mathcal{M}_{m_j,\lambda}= \bra{\mathcal{P}(P')} J^\mu(0) \ket{\mathcal{V} (P,m_j)}\epsilon_{\mu,\lambda}^*(q) |_{Q^2=0}\;,
\end{align}
where $\epsilon_{\mu,\lambda}$ is the polarization vector of the final-state photon with its spin projection $\lambda=\pm 1$.
The decay width of $\mathcal{V} \to \mathcal{P}+\gamma$ follows by averaging over the initial polarization and summing over the final
polarization. In the rest frame of the initial particle, it reads,
\begin{align}\label{eq:VPwidth}
  \begin{split}
    \Gamma(\mathcal{V} \to   \mathcal{P}+\gamma)
    =&\int\diff \Omega_q
    \frac{1}{32\pi^2}
    \frac{|\vec{q}|}{m_{\mathcal{V} }^2}
    \frac{1}{2 J_{\mathcal{V} }+1}
    \sum_{m_j,\lambda}|\mathcal{M}_{m_j,\lambda}|^2
    =
    \frac{ {(m_{\mathcal{V} }^2-m_{\mathcal{P}}^2)}^3}{ {(2m_{\mathcal{V}})}^3{(m_{\mathcal{P}}+m_{\mathcal{V}
        })}^2}
    \frac{{|V(0)|}^2}{(2 J_{\mathcal{V} }+1)\pi}
    \;.
  \end{split}
\end{align}
The momentum of the final photon is determined by the energy-momentum conservation, $|\vec{q}|=( m_{\mathcal{V} }^2-
m_{\mathcal{P}}^2)/2 m_{\mathcal{V} }$. $J_{\mathcal{V} }=1$ is the spin of the initial vector meson. To calculate the width of
$\mathcal{P} \to \mathcal{V}+\gamma$, exchange $m_{\mathcal{V} }$ and $m_{\mathcal{P}}$, and replace  $J_{\mathcal{V} }$ with
$J_{\mathcal{P} }=0$ for the initial pseudoscalar in Eq.~\eqref{eq:VPwidth}.
\subsection{Light-front dynamics}\label{subsec:lf}
In principle, the Lorentz invariant function $V(Q^2)$ defined in Eq.~\eqref{eq:Vq2_def} can be extracted from any of the four sets of hadron matrix elements, $ \mu=+,-,x,y$ ($v^\pm=v^0\pm v^z$, see definitions of the light-front variables in the Appendix). However, results from different current components may be different due to violations of Lorentz symmetry
introduced by the Fock sector truncation as well as by the modeling of systems. These approximations have led to extensive discussions in the
literature~\cite{CARBONELL1998215,DEMELO1998574,BRODSKY1999239, PhysRevD.88.025036,PhysRevD.65.094043}. The ``+'' component, known
as the ``good current'', is typically used, together with the Drell-Yan frame ($q^+=0$), to avoid contributions from pair production/annihilation in vacuum. The transverse components have been shown to be consistent with the ``+'' component in the limit of zero momentum
transfer in certain theories, such as the $\phi^3$ theory~\cite{BRODSKY1999239} and the spin-0 two-fermion systems~\cite{PhysRevD.65.094043}. Another option, the ``-'' component, is known as the ``bad current'', due to its association with the zero-mode
contributions. 

Here, we present formulae for the transition form factor, for both $J^+$ and $\vec  J_\perp$, along
with different magnetic projections ($m_j=0, \pm 1$) of the vector meson. Note that when the rotational symmetry on the transverse plane is
preserved, which is usually the case, using $J^x$ or $J^y$ component or even combinations of the two are equivalent. In
particular, we use $J^R\equiv J^x+i J^y$ to carry out the calculation in the case of the transverse current. For
any transverse vector $\vec k_\perp$, which is expressed as $(k^x, k^y)$ in the Cartesian coordinate or $(k_\perp,\theta)$ in the
polar coordinate, we will write its complex form as $k^R\equiv k^x +i k^y=k_\perp e^{i \theta}$ and $k^L\equiv k^x -i k^y=k_\perp e^{-i\theta}$. From the vector decomposition in Eq.~\eqref{eq:Vq2_def},
\begin{align}\label{eq:Jpl}
I^+_{m_j}  =&\frac{2 V(Q^2)}{m_{\mathcal{P}}+m_{\mathcal{V}}}
     \begin{dcases}
       0,  &m_j=0\\
 \frac{i}{\sqrt{2}}P^+\Delta^R,
 &m_j= 1\\
 -\frac{i}{\sqrt{2}}P^+\Delta^L,      
&m_j= -1
     \end{dcases}\\\label{eq:JR}
I^R_{m_j} 
  =&\frac{2 V(Q^2)}{m_{\mathcal{P}}+m_{\mathcal{V}}}
     \begin{dcases}
-im_{\mathcal{V}} \Delta^R,
&m_j=0\\
\frac{i}{\sqrt{2}}P^R\Delta^R, 
&m_j= 1\\       
\frac{i}{\sqrt{2}z}(z^2 m_{\mathcal{V}}^2-m_{\mathcal{P}}^2-{P'}^R\Delta^L )       , 
&m_j= -1
     \end{dcases}
\end{align}
where we have introduced two variables
\(z\equiv {P'}^+/P^+\) and \(\vec \Delta_\perp=\vec P'_\perp-z\vec P_\perp\), which are invariant under the transverse Lorentz boost specified by the velocity vector $\vec \beta_\perp$,
\begin{align}\label{eq:tr_boost}
  v^+ \to v^+,\quad \vec{v}_\perp \to \vec{v}_\perp +v^+\vec{\beta}_\perp \;.
\end{align}
This boost is kinematic and survives the Fock space truncation, whereas the full Lorentz transformation does not. The two sets of hadron matrix elements in Eqs.~\eqref{eq:Jpl} and~\eqref{eq:JR} can be related through such a boost, 
\begin{align}\label{eq:trLB}
  \begin{split}
    \bra{\mathcal{P}({P'}^+,\vec{P}'_\perp+{P'}^+\vec{\beta}_\perp)}& \vec J_\perp\ket{\mathcal{V} (P^+,
      \vec{P}_\perp+P^+\vec{\beta}_\perp, m_j)}\\
    =\bra{\mathcal{P}({P'}^+,\vec{P}'_\perp)}  & \vec J_\perp \ket{\mathcal{V} (P^+,\vec{P}_\perp,m_j)}
    +\vec{\beta}_\perp \bra{\mathcal{P}({P'}^+,\vec{P}'_\perp)} J^+\ket{\mathcal{V} (P^+,\vec{P}_\perp,m_j)} \;.
  \end{split}
\end{align}
By applying the above relation to Eqs.~\eqref{eq:Jpl} and ~\eqref{eq:JR},  we find that for $m_j=\pm 1$, $J^+$ and $J^R$ should give
the same $V(Q^2)$. For $m_j=0$, on the other hand, $V(Q^2)$ cannot be extracted from $J^+$, but can be extracted from transverse currents, such as $J^R$. 
\begin{align}
&\begin{aligned}\label{eq:Vpl_mj}
V^+_{m_j=\pm 1}(Q^2)=&
    \pm i\frac{m_{\mathcal{P}}+m_{\mathcal{V}}}{\sqrt{2}P^+\Delta^{R/L}}I^+_{\pm 1},
\end{aligned}\\
&\begin{aligned}\label{eq:VR_mj}
V^R_{m_j=1}(Q^2)=&
     i\frac{m_{\mathcal{P}}+m_{\mathcal{V}}}{\sqrt{2}P^R\Delta^R}I^R_{1},
\qquad
V^R_{m_j=-1}(Q^2)=
     i\frac{(m_{\mathcal{P}}+m_{\mathcal{V}})z}{\sqrt{2}(z^2 m_{\mathcal{V}}^2-m_{\mathcal{P}}^2-{P'}^R\Delta^L )}I^R_{-1},\\
V^R_{m_j=0}(Q^2)=&
    -i\frac{m_{\mathcal{P}}+m_{\mathcal{V}}}{2 m_{\mathcal{V}}\Delta^R}I^R_0
    \;.
\end{aligned}
\end{align}
Note that for the purpose of comparison, we label the transition form factors with their corresponding current components and the
$m_j$ values of the vector wavefunctions. In practice, the different prescriptions of extracting the same transition form factor
could provide a test of violation of the Lorentz symmetry in the calculation. In the covariant light-front dynamics, the transition form factor
is extracted from combinations of several hadron matrix elements~\cite{CARBONELL1998215}.

\subsection{Impulse approximation}
\begin{figure}[h!]
  \centering
  \includegraphics[width=.8\textwidth]{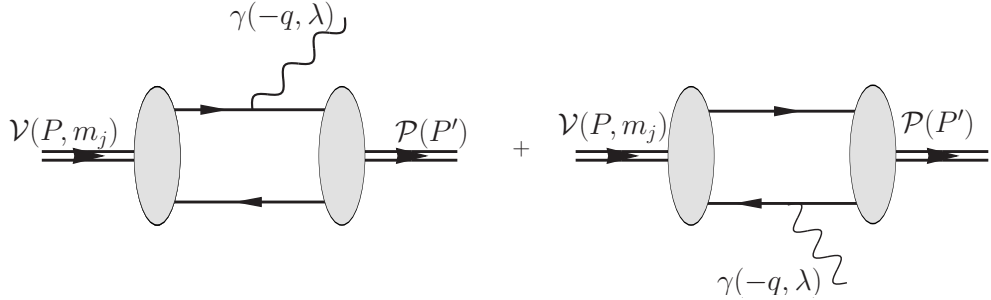}
  \caption{ Radiative transition from vector to pseudoscalar meson in $\ket{q\bar{q}}$ Fock space representation within the impulse approximation. In these
    figures light-front time $x^+$ flows to the right. The double-lines represent the hadrons.  The solid lines represent the quark or the antiquark. The wavy lines represent the probing photon. The shaded areas represent the LFWFs. 
  }\label{fig:VP_qqbar}
\end{figure}

In the impulse approximation, the interaction of the external current with the meson is the summation of its coupling to the quark and to the antiquark, as illustrated in Fig.~\ref{fig:VP_qqbar}. The vertex dressing as well as pair creation/annihilation from higher order diagrams are neglected.

The hadron matrix element can be written accordingly as a sum of the quark term and the antiquark term:
\begin{align}
 \bra{\mathcal{P}(P')} J^\mu(0)\ket{\mathcal{V} (P,m_j)}
  =e \mathcal{Q}_f \bra{\mathcal{P}(P')} J_{q}^\mu(0)\ket{\mathcal{V} (P,m_j)}
  -e \mathcal{Q}_f\bra{\mathcal{P}(P')} J_{\bar{q}}^\mu(0)\ket{\mathcal{V} (P,m_j)}
  \;.
\end{align}
The current operator is defined as $J^\mu(x) = e \sum_f Q_f \overline \psi_f(x) \gamma^\mu \psi_f(x)$ where $\psi_f(x)$ is the quark field
operator with flavor $f$ ($f=u,d,s,c,b,t$).  $J_q$ and $J_{\bar q}$ are the normal ordered pure quark ($b^\dagger b$) and antiquark ($d^\dagger d$) part of $J^\mu$, 
respectively, where $b$ ($d$) is the quark (antiquark) annihilation operator.
 The dimensionless
fractional charge of the quark is, $\mathcal{Q}_f=\mathcal{Q}_c=+2/3$ for the charm quark and $\mathcal{Q}_f=\mathcal{Q}_b=-1/3$ for the bottom
quark. The electric charge $e=\sqrt{4\pi \alpha_{\text{EM}}}$. For quarkonium, due to the charge conjugation symmetry, the antiquark gives the
same contribution as the quark to the total hadronic current. So, for our purpose, we calculate the hadron matrix element for the quark part. As such, we
compute $\hat{V}(Q^2)$ which is related to $V(Q^2)$ by
\[
  V(Q^2)=2e \mathcal{Q}_f \hat{V}(Q^2)\;.
\]

The hadron matrix element can be written explicitly in terms of the convolution of LFWFs. To begin with, the valence Fock space representation of quarkonium reads:
\begin{align}
  \begin{split}
  \ket{\psi_h(P, j, m_j)} = &
  \sum_{s, \bar{s}}\int_0^1\frac{\diff x}{2x(1-x)} \int \frac{\diff^2 \vec k_\perp}{{(2\pi)}^3}
  \, \psi^{(m_j)}_{s\bar s/h}(\vec k_\perp, x) \\
 & \times \frac{1}{\sqrt{N_c}}\sum_{c=1}^{N_c} b^\dagger_{sc}\big(xP^+, \vec k_\perp+x\vec P_\perp\big) d^\dagger_{\bar s c}\big((1-x)P^+,
  -\vec k_\perp+(1-x)\vec P_\perp\big) \ket{0},
\end{split}
\end{align}
where the color index $c=1,2,\ldots, N_c$, and the number of quark colors $N_c=3$. $\psi^{(m_j)}_{s\bar s/h}(\vec k_\perp, x)$ is the LFWF written in relative coordinates. $x\equiv p^+/P^+$ is the
longitudinal light-front momentum fraction, $\vec k_\perp=\vec p_\perp-x\vec P_\perp$ is the relative transverse momentum, where
$p$ is the single-particle 4-momentum of the quark. $s$ represents the fermion spin projection in the $x^-$ direction. 

The hadron matrix element in the Drell-Yan frame ($q^+=0$, i.e. $z=1$) follows,
\begin{align}\label{eq:Jq}
  \begin{split}
    \bra{\mathcal{P}(P')} J_q^\mu(0)\ket{\mathcal{V} (P,m_j)}
    =
    &  \sum_{s,\bar s, s'}
    \int_0^1\frac{\diff x}{2x^2(1-x)}
    \int\frac{\diff^2\vec{k}_\perp}{{(2\pi)}^3}
    \psi_{s \bar s/\mathcal{V}}^{(m_j)}(\vec{k}_\perp, x)
    \psi_{s'\bar s/\mathcal{P}}^{ *}
    (\vec{k}_\perp+(1-x)\vec{q}_\perp , x)  \\
    &
    \bar{u}_{s' }(xP^+ , \vec{k}_\perp+x\vec{P}_\perp+\vec{q}_\perp)
    \gamma^\mu u_{s}(xP^+, \vec{k}_\perp +x\vec{P}_\perp)
    \;.
  \end{split}
\end{align}

We could then obtain the transition form factor from such hadron matrix elements according to Eqs.~\eqref{eq:Vpl_mj} and ~\eqref{eq:VR_mj}. 
Ideally, $\hat V(Q^2)$ is independent of the spin projection $m_j$ and the current components. Nevertheless, one needs to carefully choose the proper matrix elements to evaluate certain quantities, when approximations break the Lorentz symmetry~\cite{Li:2017uug}. For instance, there are
different ways of choosing matrix elements to calculate the spin-1 electromagnetic form factors when the angular condition is
violated~\cite{Grach:1983hd}. Among those, some are preferred 
in the sense that unphysical terms could be partially or
entirely suppressed~\cite{KARMANOV1996316,PhysRevD.65.094043}. 

In the case of the M1 transition form factor $\hat V(Q^2)$, using the combination of the transverse current $J^R$ with the $m_j=0$
polarization of the vector meson, according to the expression in Eq.~\eqref{eq:VR_mj},
would  give the LFWF representation as, 
\begin{align}\label{eq:Vmj0}
\begin{split}
   \hat V_{m_j=0}(Q^2)
  =&
  \frac{i (m_{\mathcal{P}}+m_{\mathcal{V}})}{m_{\mathcal{V}}}
    \int_0^1\frac{\diff x}{2x^2(1-x)}
    \int\frac{\diff^2\vec{k}_\perp}{{(2\pi)}^3}
[
 -\frac{1}{2}\psi_{\uparrow\downarrow+\downarrow\uparrow/\mathcal{V}}^{(m_j=0)}(\vec{k}_\perp, x)
  \psi_{\uparrow\downarrow-\downarrow\uparrow/\mathcal{P}}^{ *}
    (\vec{k}_\perp+(1-x)\vec{q}_\perp , x)\\
&
    +\psi_{\downarrow\downarrow/\mathcal{V}}^{(m_j=0)}(\vec{k}_\perp, x)
  \psi_{\downarrow \downarrow\mathcal{P}}^{ *}
    (\vec{k}_\perp+(1-x)\vec{q}_\perp , x)
]
    \;,
\end{split}
\end{align}
where we define $\psi_{\uparrow\downarrow\pm \downarrow\uparrow}\equiv (\psi_{\uparrow\downarrow}\pm \psi_{\downarrow\uparrow})/\sqrt{2}$. Note that in deriving Eq.~\eqref{eq:Vmj0}, we have taken advantage of symmetries in LFWFs, $\psi_{\uparrow\downarrow-\downarrow\uparrow/\mathcal{V}}^{(m_j=0)}=0$ and $\psi_{\uparrow\downarrow+\downarrow\uparrow/\mathcal{P}}=0$.

The traditionally used good current $J^+$ is also worth looking at. As we have discussed, with this current component, the
transition form factor can be extracted only from the $m_j=\pm 1$ polarizations of the vector meson. We present the expression for
$m_j=1$ according to Eq.~\eqref{eq:Vpl_mj}, while the expression for $m_j=-1$ is similar. It is evident from this expression that the overlapped spin components of the
two wavefunctions indicate no spin-flip (between spin-triplet and spin-singlet), which may appear counter-intuitive for the M1
transition. Indeed, this calculation relies on subdominant terms and is less robust, as we will discuss in the
following section for heavy quarkonia.
\begin{align}\label{eq:Vmj1}
  \hat V_{m_j=1}(Q^2)
  =
  \frac{\sqrt{2}(m_{\mathcal{P}}+m_{\mathcal{V}})}{i q^R}
  \sum_{s, \bar s}
  \int_0^1\frac{\diff x}{2x(1-x)}
  \int\frac{\diff^2\vec{k}_\perp}{{(2\pi)}^3}
  \psi_{s \bar s/\mathcal{V}}^{(m_j=1)}(\vec{k}_\perp, x)
  \psi_{s \bar s/\mathcal{P}}^{ *}
  (\vec{k}_\perp+(1-x)\vec{q}_\perp , x)
  \;.
\end{align}

\subsection{The nonrelativistic limit}\label{sec:NR_limit}
In the nonrelativistic limit, the M1 transition with the same radial
or angular quantum numbers (e.g. $nS\to nS+\gamma$), is often referred to as \emph{allowed}, for which the transition amplitude is large and $\hat V(0)\to 2$ as a
result of the similarity between the spatial wavefunctions of the vector
and the pseudoscalar mesons with the same spatial quantum numbers; whereas the transition between states with different radial or angular excitations is referred to as \emph{hindered}, for which the transition amplitude is zero and $\hat V(0)\to 0$ at leading order due to the orthogonality of the wavefunctions~\cite{PhysRevD.64.074011,Brambilla:2005zw,Lewis:2011ti,Eichten:2007qx}. 
The deviations of experimentally measured results from those nonrelativistic limits
indicate relativistic effects~\cite{PDG2016}. For a heavy quarkonium system, which is close to the nonrelativistic
domain, such deviations are expected to be small but nonzero.

\begin{figure}[!htp]
  \centering 
  \subfloat[\ $\mathcal{P}(nS)$~\label{fig:DS_Pn}]{
            \includegraphics[width=0.3\textwidth]{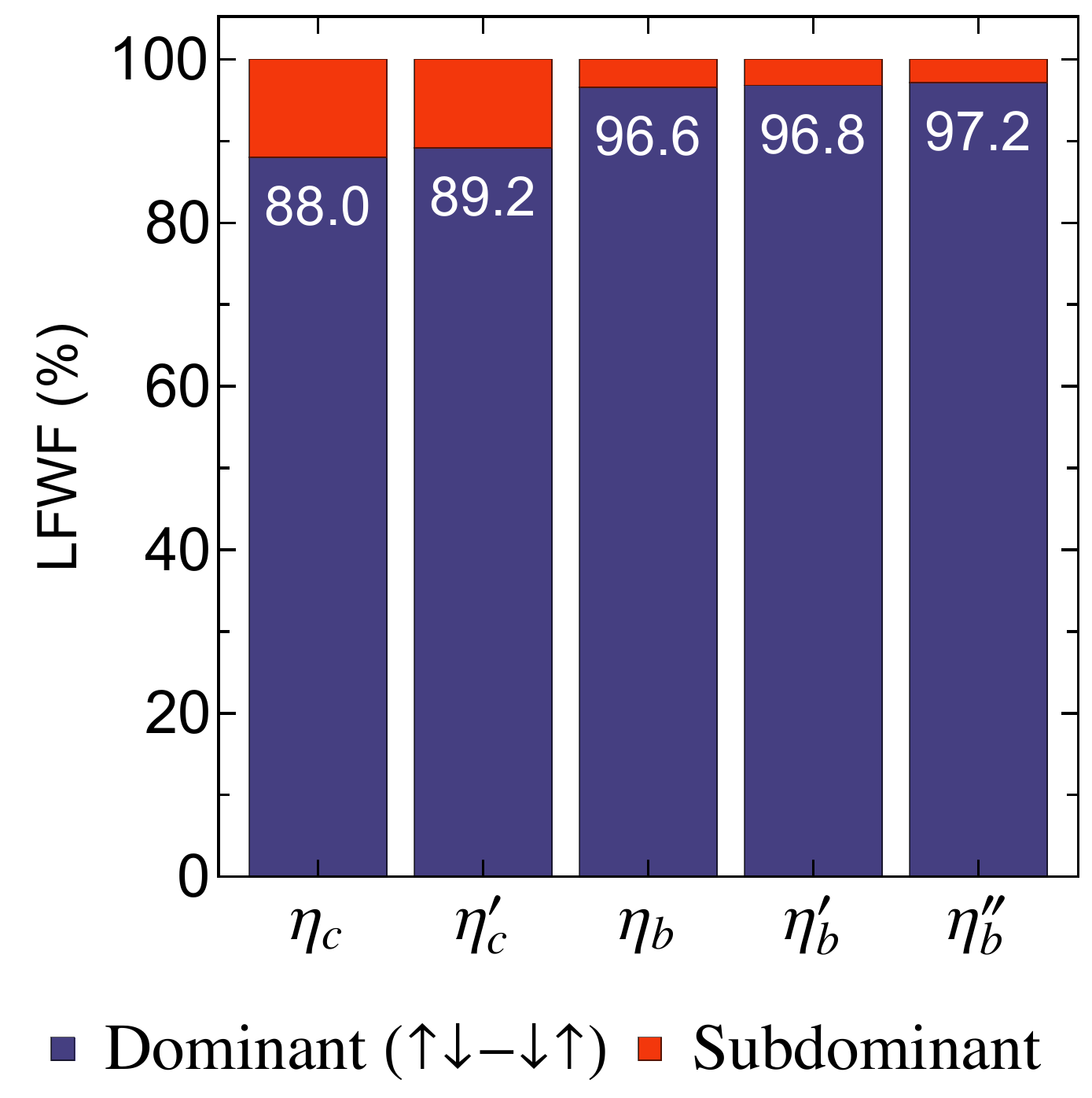}
  }\quad
  \subfloat[\ $\mathcal{V}^{(m_j=0)}(nS)$~\label{fig:DS_VnS_mj0}]{
    \includegraphics[width=0.3\textwidth]{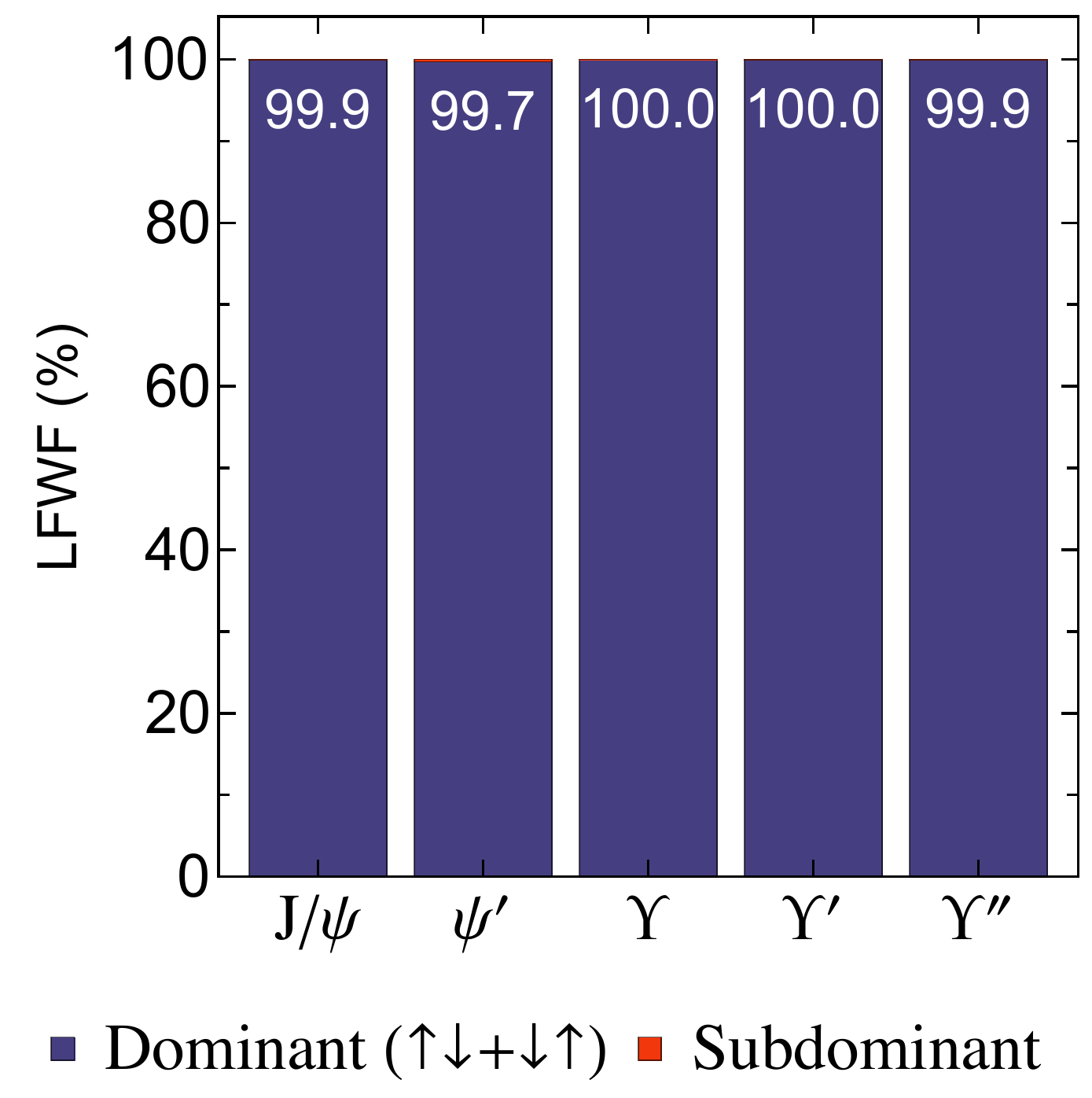}
  }
\quad
  \subfloat[\ $\mathcal{V}^{(m_j=1)}(nS)$~\label{fig:DS_VnS_mj1}]{
    \includegraphics[width=0.3\textwidth]{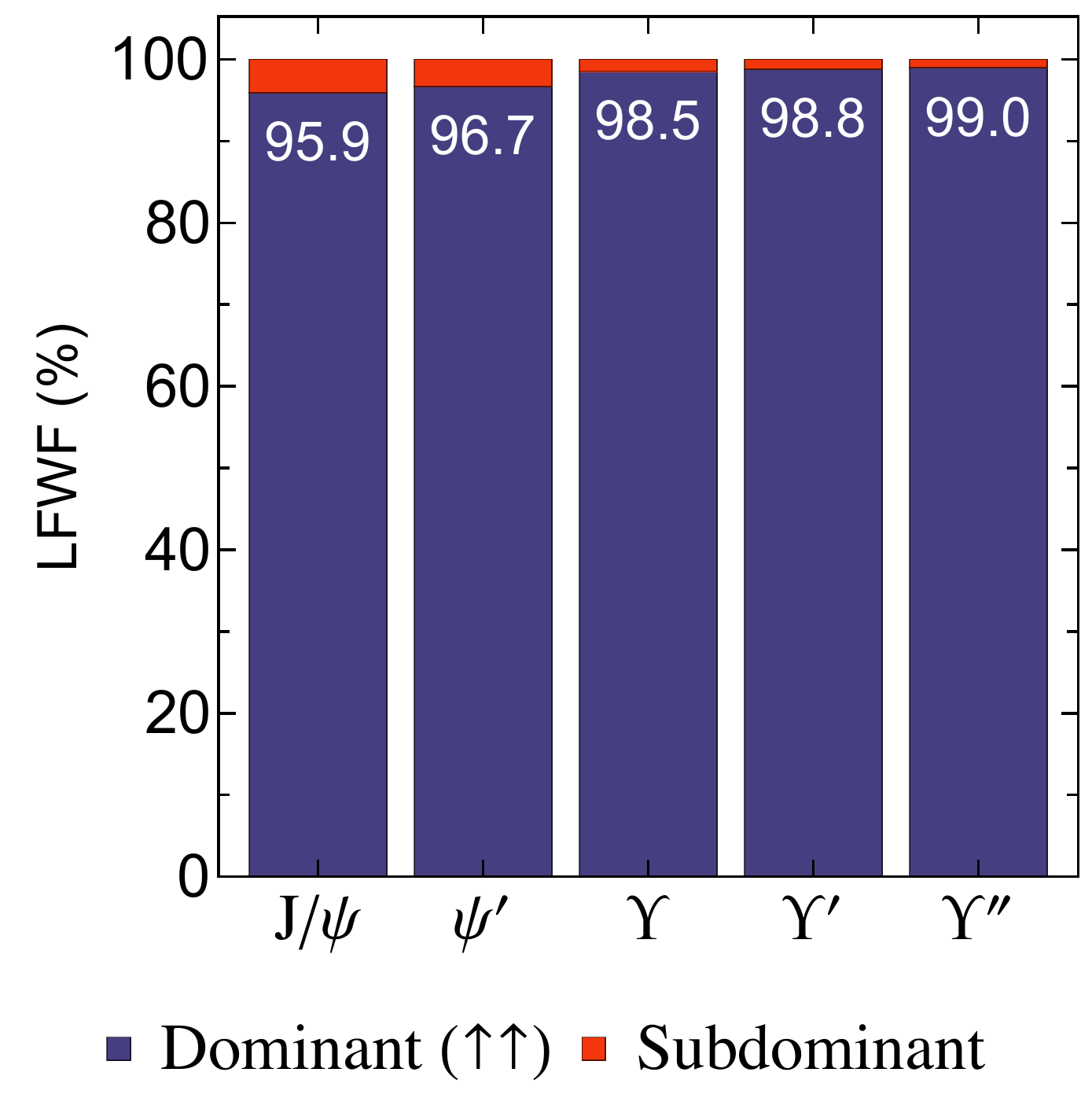}
  }  \\
  \subfloat[\ $\mathcal{V}^{(m_j=0)}(n^3D_1)$~\label{fig:DS_VnD_mj0}]{
    \includegraphics[width=0.38\textwidth]{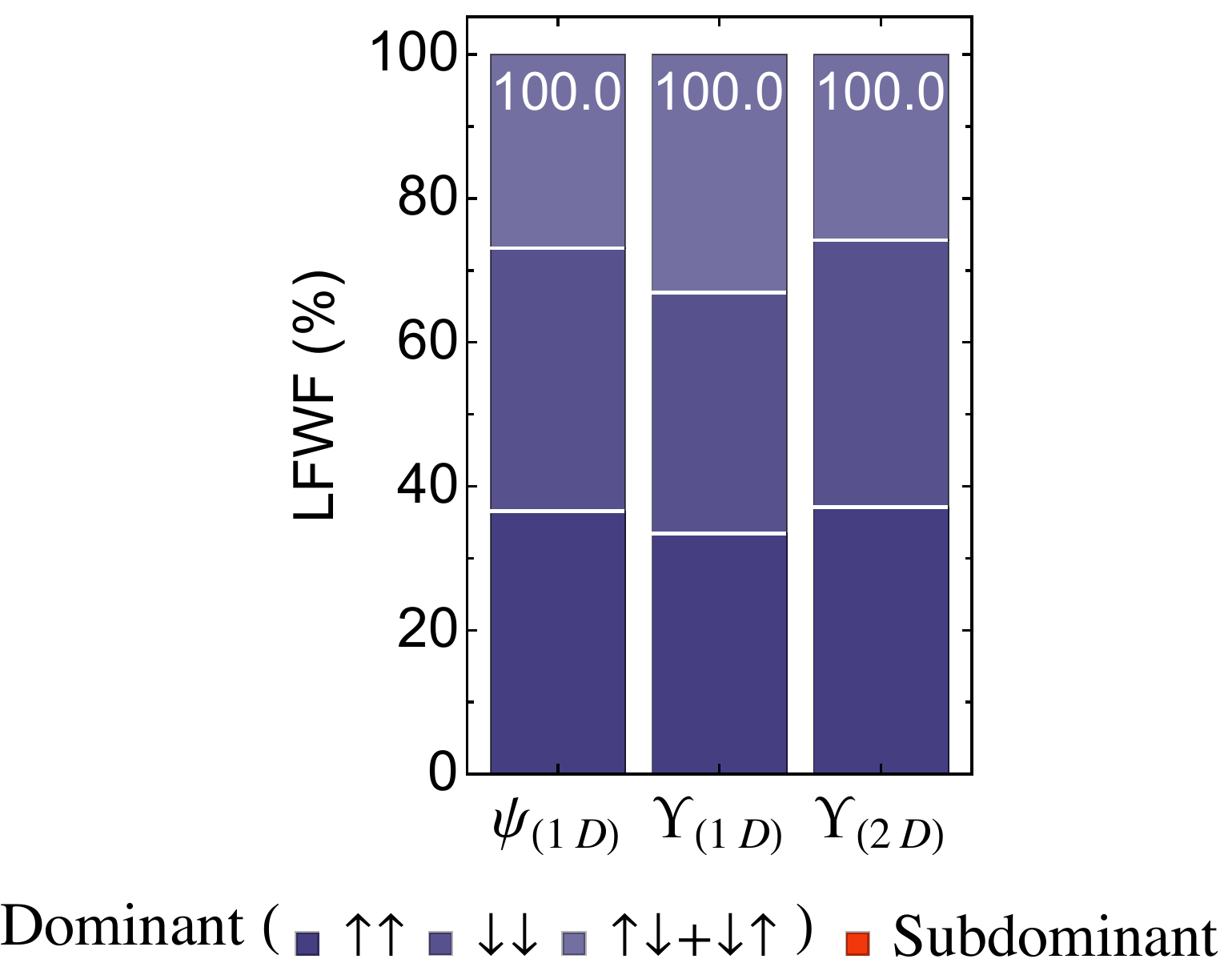}
  }\quad
  \subfloat[\ $\mathcal{V}^{(m_j=1)}(n^3D_1)$~\label{fig:DS_VnD_mj1}]{
    \includegraphics[width=0.38\textwidth]{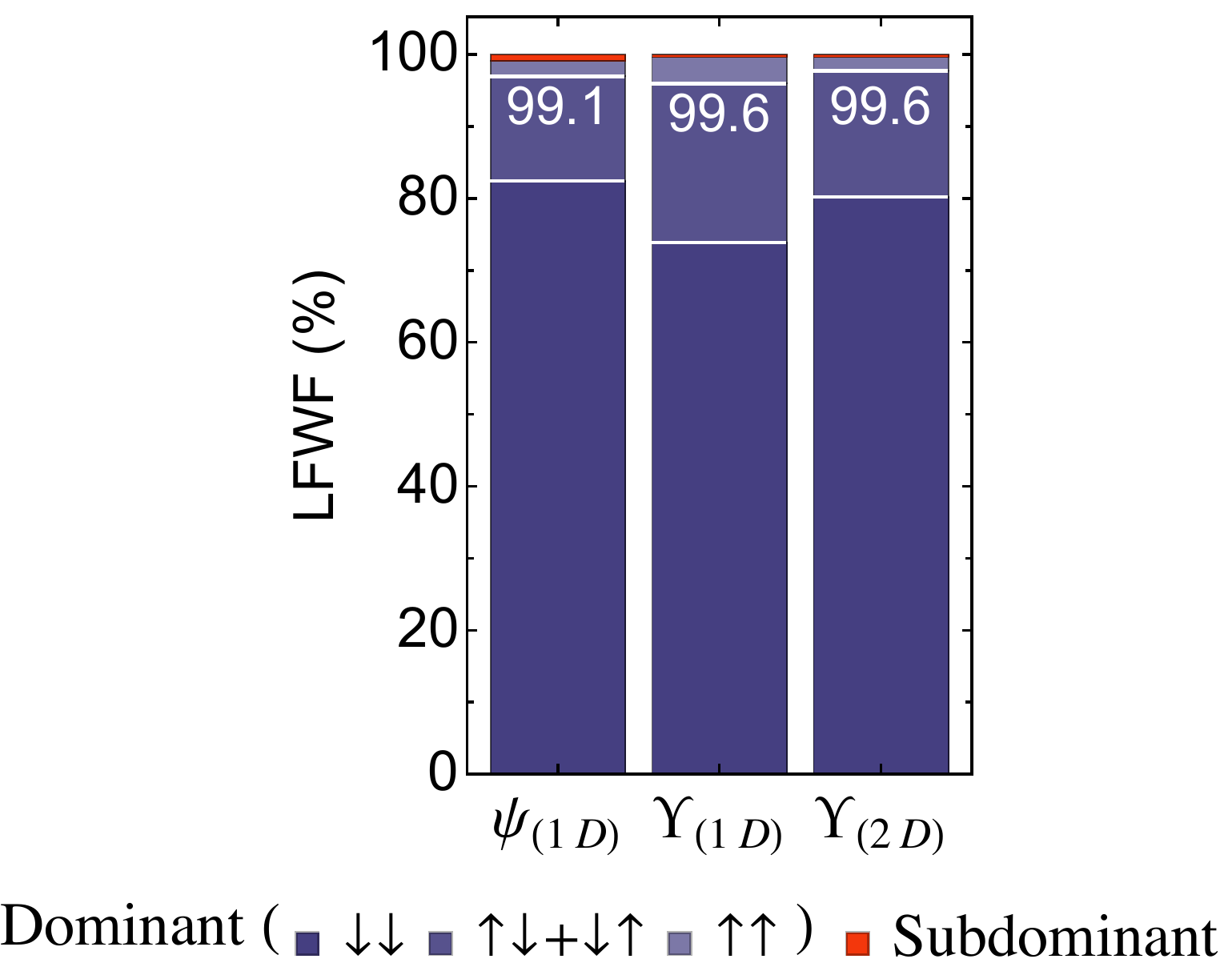}
  }
  \caption{ Comparison of dominant and subdominant LFWF components for pseudoscalar and vector in heavy quarkonia. Single (double) apostrophe stands for the radial excited 2S (3S) state. LFWFs are taken from the $N_{\max}=L_{\max}=32$ result of Ref.~\cite{Yang_run}. The numbers in white suggest the occupancy of the dominant spin components for each state.}
  \label{fig:DS}
\end{figure}

The wavefunctions of heavy quarkonia, treated as relativistic bound states,  are dominated by those components
that are non-vanishing and reduce to the nonrelativistic wavefunction in the nonrelativistic limit. These wavefunction components
are therefore referred to as the \emph{dominant} components. It is necessary to emphasize that despite the correspondence between the dominant spin components and the nonrelativistic wavefunctions, the former carries relativistic contributions when solved in a relativistic formalism. There are also wavefunction components of purely relativistic origin, which vanish in the nonrelativistic limit and
are therefore \emph{subdominant}. In practice, the dominant components tend to be better constrained, while the subdominant ones are more
sensitive to the model and numerical uncertainties.
For the pseudoscalar states resembling S-waves (in particular $n^1S_0$), $\eta_c(nS)$ and $\eta_b(nS)$, their dominant components are the spin-singlets $\psi_{\uparrow\downarrow-\downarrow\uparrow/\mathcal{P}}$, while relativistic treatments would also allow them to have subdominant components, such as $\psi_{\uparrow\uparrow/\mathcal{P}}=\psi^*_{\downarrow\downarrow/\mathcal{P}}$.
Analogously, for the vector states of heavy quarkonia resembling S-waves (in particular $n^3S_1$), $\psi(nS)$ and $\Upsilon(nS)$,
the dominant components are the spin-triplets,
$\psi^{m_j=0}_{\uparrow\downarrow+\downarrow\uparrow/\mathcal{V}}$, $\psi^{m_j=1}_{\uparrow\uparrow/\mathcal{V}}$ and
$\psi^{m_j=-1}_{\downarrow\downarrow/\mathcal{V}}$.
For those vector states identified as D-waves, $\psi(n^3D_1)$ and $\Upsilon(n^3D_1)$, where orbital excitations occur, all the spin-triplet components
$\psi^{m_j=0,\pm 1}_{\uparrow\downarrow+\downarrow\uparrow/\mathcal{V}}$, $\psi^{m_j=0, \pm 1}_{\uparrow\uparrow/\mathcal{V}}$
and $\psi^{m_j=0,\pm 1}_{\downarrow\downarrow/\mathcal{V}}$ exist in the nonrelativistic limit and are considered dominant, and only the spin-singlet components $\psi^{m_j=0,\pm
  1}_{\uparrow\downarrow-\downarrow\uparrow/\mathcal{V}}$ are subdominant.
In detail, the spin components with larger orbital angular momentum projection $m_\ell=m_j-s-\bar s$, 
$\psi^{m_j=0}_{\uparrow\uparrow/\mathcal{V}}=-\psi^{* m_j=0}_{\downarrow\downarrow/\mathcal{V}} (m_\ell=\pm 1)$ and
$\psi^{m_j=1}_{\downarrow\downarrow/\mathcal{V}} (m_\ell=2)$, have the largest occupancy.
The less occupied components,  $\psi^{m_j=0}_{\uparrow\downarrow+\downarrow\uparrow/\mathcal{V}} (m_\ell=0)$, $\psi^{m_j=1}_{\uparrow\downarrow+\downarrow\uparrow/\mathcal{V}} (m_\ell=1)$ and $\psi^{m_j=1}_{\uparrow\uparrow/\mathcal{V}} (m_\ell=0)$, could also exist in the nonrelativistic limit.  
Moreover, the spin components with $m_\ell=0$
admit the admixtures of S-waves, though the actual amount of such admixtures is small and sensitive to both the model parameters and the truncation.
For example, the $\psi(3770)$ [$\psi(1D)$] state, though primarily a $1^3D_1$ state, has contributions from $n^3S_1$
states (notably $2^3S_1$)~\cite{Richard:1979fc,Rosner:2004wy,Eichten:2004uh,Eichten:2005ga}, and these S-wave admixtures are responsible for the $nD \to n'S+\gamma$ transitions~\cite{Eichten:1978tg,Kwong:1988ae,Rosner:2001nm,Eichten:2007qx}.  
In order to have a more intuitive view of the dominant and subdominant spin components for those states, we take the LFWFs from
Ref.~\cite{Yang_run} to show in Fig.~\ref{fig:DS} the proportions of those dominant and subdominant components of heavy quarkonia. For all those pseudoscalar and vector states, the dominant terms could each occupy $88\% \sim 100\%$ of the whole LFWF,
suggesting that the heavy quarkonium indeed resembles a nonrelativistic system. The comparison between the same states of the charmonium
and those of the bottomonium also reveals that the dominant component is more pronounced in the heavier, and less relativistic,
system. 

It follows that in calculating the transition form factors, we can examine the two procedures, $\hat V_{m_j=0}(Q^2)$ presented in Eq.~\eqref{eq:Vmj0} and $\hat V_{m_j=1}(Q^2)$ presented in Eq.~\eqref{eq:Vmj1}, in terms of their 
proximities to the nonrelativistic domain. 
The result of $\hat V_{m_j=0}(Q^2)$ mainly comes from the overlap of the dominant components, $\psi_{\uparrow\downarrow+\downarrow\uparrow/\mathcal{V}}^{(m_j=0)}
  \psi_{\uparrow\downarrow-\downarrow\uparrow/\mathcal{P}}^{ *}$, whereas even the major part of $\hat V_{m_j=1}(Q^2)$ involves the subdominant components, such as 
 $\psi_{\uparrow\uparrow/\mathcal{V}}^{(m_j=1)}\psi_{\uparrow \uparrow/\mathcal{P}}^{ *}$ and  $\psi_{\uparrow\downarrow-\downarrow\uparrow/\mathcal{V}}^{(m_j=1)}\psi_{\uparrow\downarrow-\downarrow\uparrow/\mathcal{P}}^{ *}$.
In heavy quarkonium, the dominant components tend to be better constrained than the subdominant ones which suggests that $\hat V_{m_j=0}(Q^2)$ is more robust than $\hat V_{m_j=1}(Q^2)$. 

\begin{figure}[!htp]
  \centering 
  \subfloat[\ $J/\psi(1S)\to \eta_c (1S)+\gamma$~\label{fig:1s1s}]{
    \includegraphics[width=0.47\textwidth]{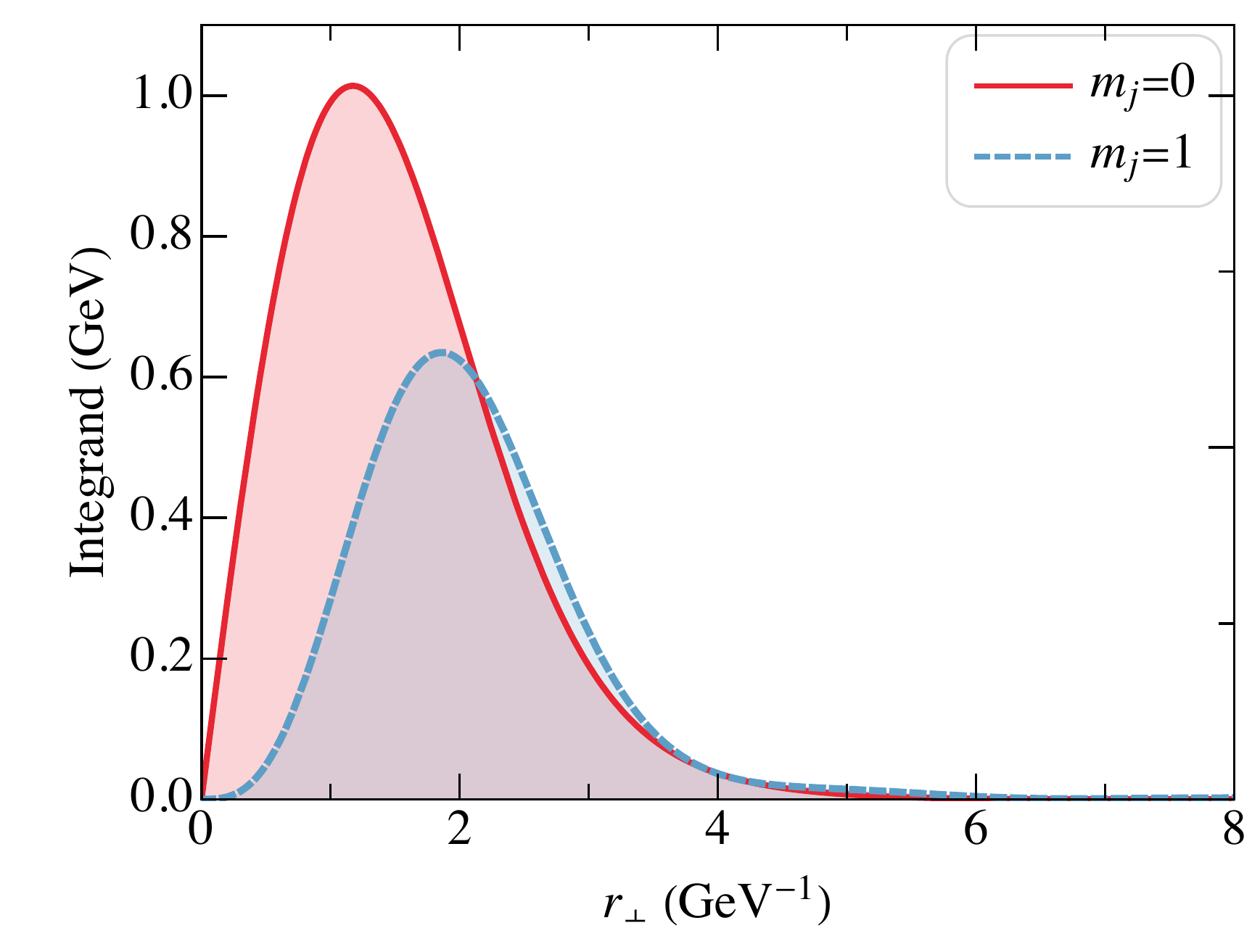}
  }\quad
  \subfloat[\ $ \eta_c (2S)\to J/\psi(1S)+\gamma$~\label{fig:1s2s}]{
    \includegraphics[width=0.47\textwidth]{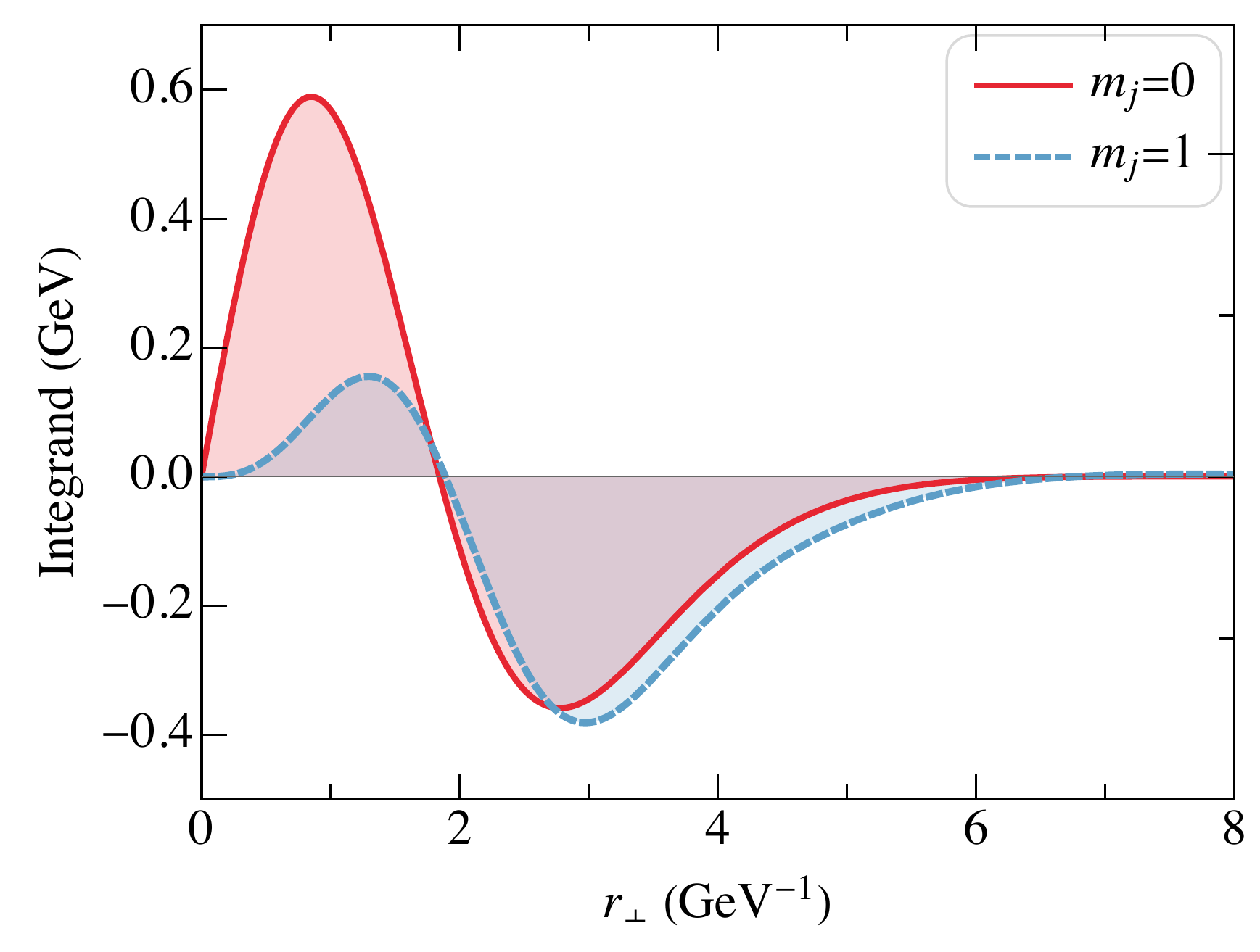}
  }
  \caption{Integrands of $\hat V(0)$ according to Eqs.~\eqref{eq:ovlp_JR} ($m_j=0$) and~\eqref{eq:ovlp_Jpl} ($m_j=1$). As a representative of the allowed ($nS\to nS+\gamma$) transitions, the integrand in (a) has the same sign in the
    entire $r_\perp$ region. On the other hand, (b) involves a transition
    with radial excitation, which is sensitive to small changes in the cancellations between positive and negative contributions.}
  \label{fig:overlaps}
\end{figure}

The nonrelativistic limit can be achieved for $\hat V_{m_j=0}(Q^2)$ by adopting nonrelativistic wavefunctions where only the dominant spin components exist. However, with $\hat V_{m_j=1}(Q^2)$,
simply taking the nonrelativistic wavefunction would always lead to zero since the expression in Eq.~\eqref{eq:Vmj1} involves the vanishing subdominant terms. 
To be specific, we examine the transition form factors at $Q^2=0$, where they can be interpreted as the overlaps of wavefunctions
in coordinate space [$\tilde{ \psi}_{s \bar s}^{(m_j)}(\vec r_\perp, x)$], shown in Eqs.~\eqref{eq:ovlp_JR} and~\eqref{eq:ovlp_Jpl}. Though equivalent to Eqs.~\eqref{eq:Vmj0}
and~\eqref{eq:Vmj1} at $Q^2=0$ respectively, Eqs.~\eqref{eq:ovlp_JR} and~\eqref{eq:ovlp_Jpl} do not have the troubling factor of $1/q^R$, and
are therefore more intuitive for the purpose of illustration.
\begin{align}
&\begin{aligned}\label{eq:ovlp_JR}
\hat V_{m_j=0}(0)
  =  \int_0^{\infty}  \diff  r_\perp
      \bigg\{&   \frac{m_{\mathcal{P}}+m_{\mathcal{V}}}{4\pi m_{\mathcal{V}} }
  \int_0^1 \diff x
      \int_0^{2\pi}\diff \theta\ \frac{r_\perp}{x}\\
&\times[
-\frac{1}{2} 
\tilde{ \psi}_{\uparrow \downarrow+\downarrow\uparrow/\mathcal{V}}^{(m_j=0)}   (r_\perp, \theta , x)\tilde{\psi}_{\uparrow \downarrow-\downarrow\uparrow/ \mathcal{P}}^*(r_\perp, \theta , x)
+\tilde{ \psi}_{\downarrow \downarrow/\mathcal{V}}^{(m_j=0)}   (r_\perp, \theta , x)
      \tilde{\psi}_{\downarrow\downarrow/ \mathcal{P}}^*(r_\perp, \theta , x)
]
      \bigg\}
\end{aligned}\\
&\begin{aligned}\label{eq:ovlp_Jpl}
  \hat V_{m_j=1}(0)
  = \int_0^{\infty} \diff  r_\perp
      \bigg\{&  \frac{\sqrt{2}(m_{\mathcal{P}}+m_{\mathcal{V})}}{4\pi } 
  \int_0^1\diff x
  \int_0^{2\pi}\diff \theta\  (1-x) r_\perp^2  \cos\theta\\
      &\times \sum_{s, \bar s}\tilde{ \psi}_{s \bar s/\mathcal{V}}^{(m_j=1)}   (r_\perp, \theta , x)
      \tilde{\psi}_{s \bar s/\mathcal{P}}^{ *}
  (r_\perp, \theta , x)
     \bigg\}
\end{aligned}
\end{align}
Note that in the nonrelativistic limit, the wavefunctions of the respective pseudoscalar and vector states with the same radial
and angular numbers are identical in their spatial dependence, and they only differ in their spin structures. For the allowed
transition, $\hat V_{m_j=0}(Q^2=0)\to 2$ due to the normalization of the spatial wavefunctions, which can be seen from
Eq.~\eqref{eq:ovlp_JR} along with taking $x\to 1/2+k_z/(2m_q)$~\cite{Yang_run} and small hyperfine splitting
$m_{\mathcal{P}}\approx m_{\mathcal{V}}$. For the hindered transition, $\hat V_{m_j=0}(Q^2=0)\to 0$ due to the orthogonality of
the two wavefunctions. Such a nonrelativistic reduction that takes advantage of the near orthonormality of wavefunctions is
reminiscent of the nonrelativistic quark model (see Refs.~\cite{PhysRevD.64.074011,Brambilla:2005zw,Lewis:2011ti}). 
However, for $\hat V_{m_j=1}(Q^2)$, the realization of the nonrelativistic limits depends strongly on the details of the subdominant wavefunctions which are less constrained in the parameter fitting. For the hindered transition, where cancellation occurs, this leads to a strong sensitivity to the model parameter and potentially to the truncation.
Fig.~\ref{fig:overlaps} presents the integrands (those inside $\{\ldots\}$) of $\hat V_{m_j=0}(0)$ and $\hat V_{m_j=1}(0)$ according to Eqs.~\eqref{eq:ovlp_JR} and~\eqref{eq:ovlp_Jpl} for an allowed ($1S\to
1S+\gamma$) as well as a hindered ($2S\to 1S+\gamma$) transition. In the left panel of Fig.~\ref{fig:overlaps}, the integrands of the allowed transition have no nodes resulting from the coherent overlaps of the two wavefunctions. 
 On the other hand, the right panel of Fig.~\ref{fig:overlaps} shows significant cancellations of contributions from the integrands which change sign due to nodes in the $2S$ wavefunctions.

Based on these lines of reasoning, we take $\hat V(Q^2)=\hat V_{m_j=0}(Q^2)$, using the transverse current, to evaluate the transition form factors for heavy quarkonia. The less robust $\hat V_{m_j=1}(Q^2)$, using the plus current, which has strong sensitivity to the violation of rotational symmetry
will also be presented for comparison.
\section{ Calculation in Basis Light-front Quantization}\label{sec:BLFQ}
We adopt wavefunctions of heavy quarkonia from recent work~\cite{Yang_fix, Yang_run} in the BLFQ approach~\cite{1stBLFQ}. The effective Hamiltonian extends the holographic QCD~\cite{holography}
by introducing the one-gluon exchange interaction with a running coupling~\cite{positronium}. The starting point for the model of Refs.~\cite{Yang_fix, Yang_run} is transverse light-front
holography, inspired by string theory, which approximates QCD at long distance. As a complementary part, the longitudinal
confining potential is introduced to allow a more consistent treatment of the mass term and the longitudinal excitation. The one-gluon exchange
implements the short-distance physics and 
determines the spin structure of the mesons. The mass spectrum and LFWFs are the direct solutions of the eigenvalue equation, and
are obtained by diagonalizing the Hamiltonian in a basis representation. The spectrum agrees with the PDG data with an r.m.s mass
deviation of  30 to 40 MeV for states below the open flavor thresholds. The LFWFs have been used to produce several observables and are in reasonable agreement with experiments and other theoretical approaches~\cite{sofia}. We now use these same LFWFs
to calculate radiative transitions with the formalism described in Sec.~\ref{sec:TFF}. 

\begin{figure}
  \centering
    \includegraphics[width=0.48\textwidth]{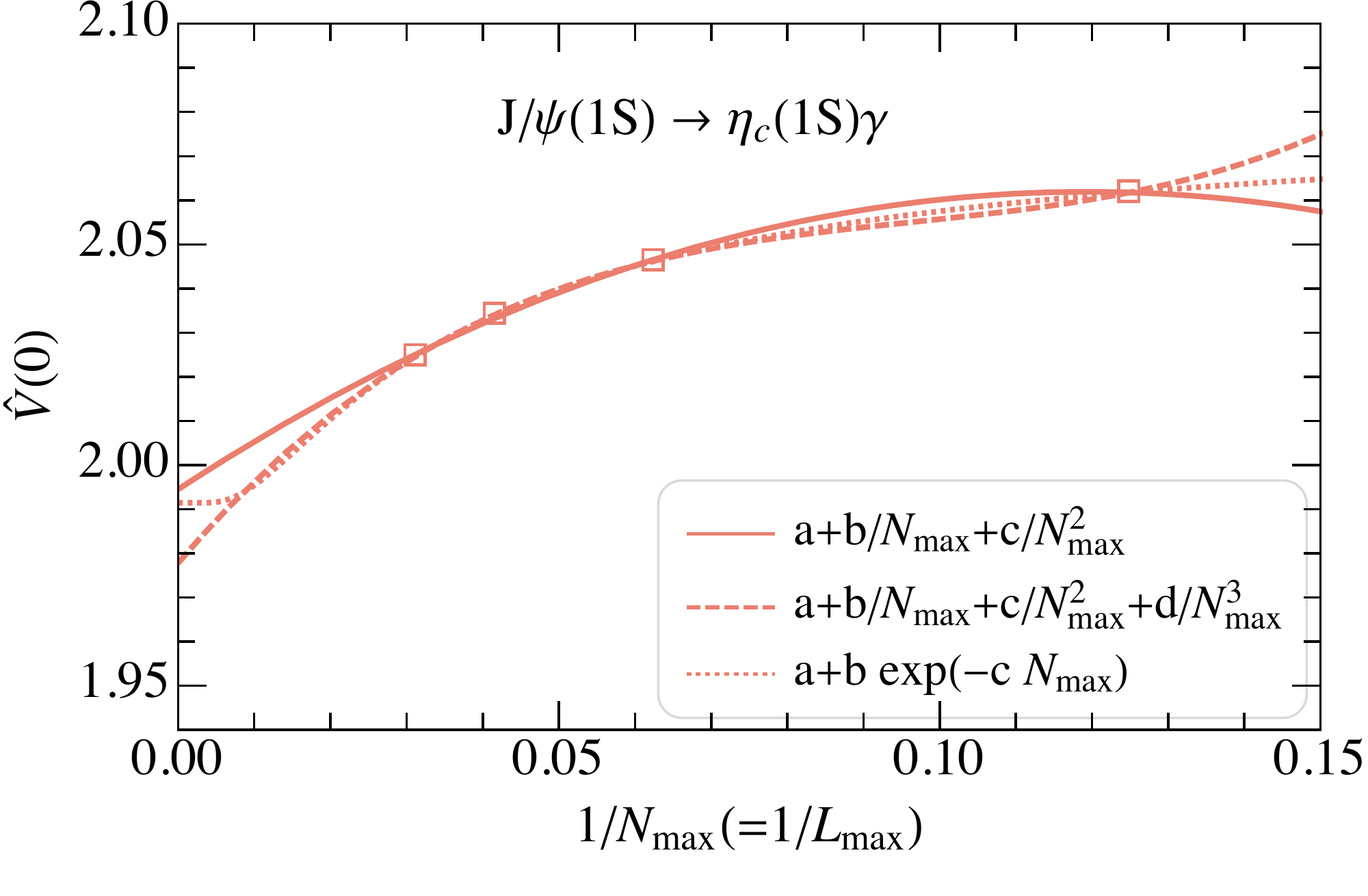}
    \includegraphics[width=0.48\textwidth]{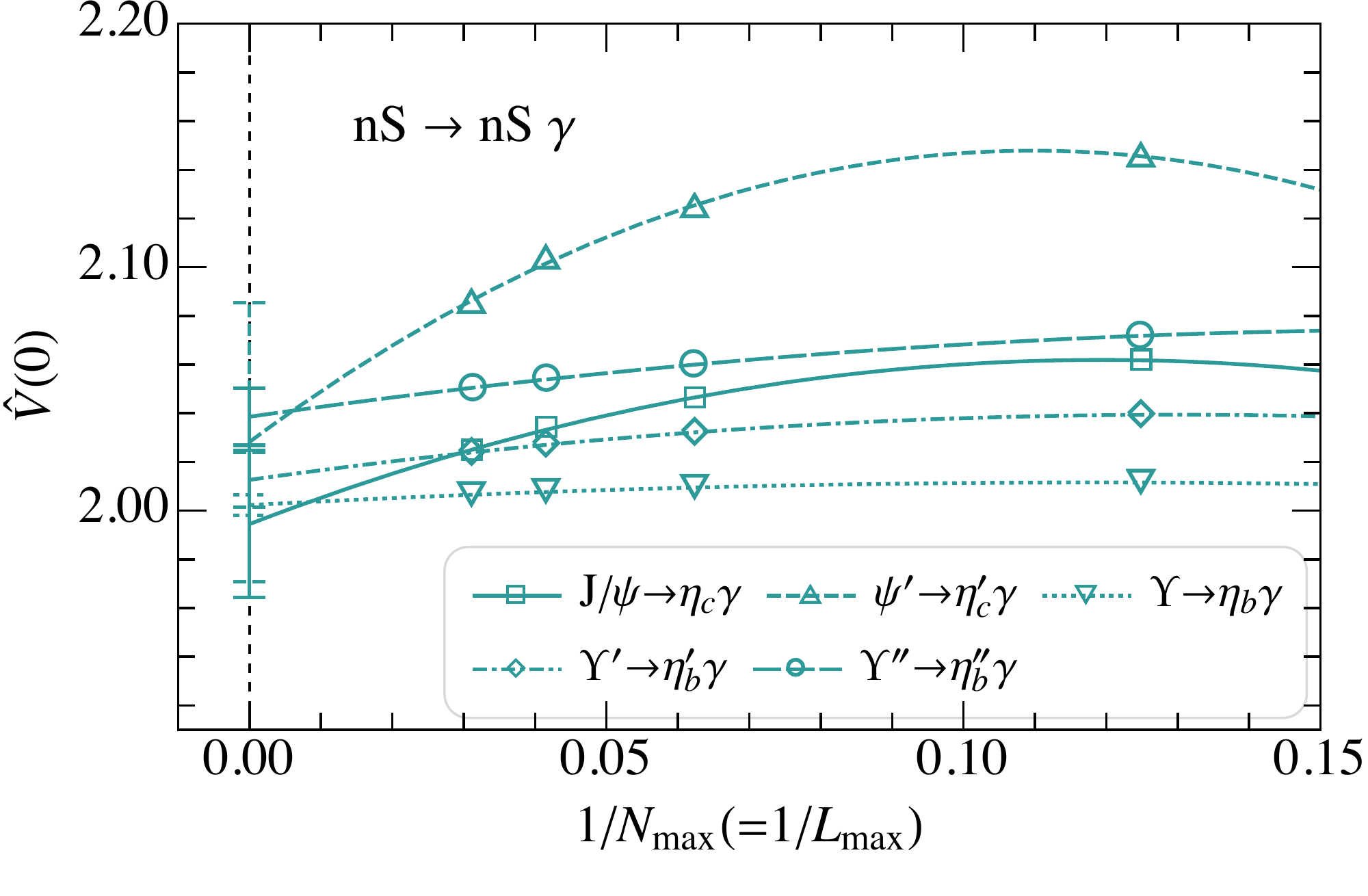}
      \caption{The $N_{\max}$ convergence of $\hat V(0)$. Transition form factors are evaluated with LFWFs at $N_{\max}=L_{\max}=8,16,24$ and $32$ according to Eq.~\eqref{eq:Vmj0}. The left panel shows the extrapolation of $\hat V(0)$ to $N_{\max}=L_{\max}\to \infty$ for $J/\psi(1S)\to \eta_c(1S)\gamma$, using three different fitting functions. We adopt the first one, $a+b/N_{\max}+c/N^2_{\max}$, to obtain the results of $\hat V(0)$ in this paper. The right panel shows the $N_{\max}(=L_{\max})$ extrapolation for all allowed transitions. The error bars indicate the difference between the result in the largest basis ($N_{\max} = L_{\max}=32$) and the extrapolated value.}
  \label{fig:extrap}
\end{figure}

The LFWFs are solved in the valence Fock sector using a basis function representation:
\begin{equation}\label{eq:LFWF}
  \psi^{(m_j)}_{s\bar s/h}(\vec k_\perp, x) = \sum_{n, m, l} \psi_h(n, m, l, s, \bar s) \, \phi_{nm}(\vec k_\perp/\sqrt{x(1-x)}) \chi_l(x).
\end{equation}
In the transverse direction, the 2D harmonic oscillator (HO) function $\phi_{nm}$ is adopted as the basis. In the longitudinal
direction, we use the modified Jacobi polynomial $\chi_l$ as the basis. $m$ is the orbital angular momentum projection, related to the
total angular momentum projection as $m_j=m+s+\bar s$, which is conserved by the Hamiltonian. The basis space is truncated by
their reference energies in dimensionless units:
\begin{align}
  2n+|m|+1\le N_{\max},\qquad 0\le l \le L_{\max}.
\end{align}
As such, the $N_{\max}$-truncation provides a natural pair of UV and IR cutoffs: $\Lambda_{\textsc{uv}} \simeq \kappa\sqrt{N_{\max}}$,
$\lambda_{\textsc{ir}} \simeq \kappa/\sqrt{N_{\max}}$, where $\kappa$ is the oscillator basis energy scale parameter. $L_{\max}$ represents the
resolution of the basis in the longitudinal direction. See Ref.~\cite{Yang_run} for details on basis functions and parameter values. The LFWFs are calculated at $N_{\max}=L_{\max}=8,16,24$ and $32$. 
Transition form factors are computed at each of these basis truncations. Fig.~\ref{fig:extrap} shows the convergence trends of $\hat V(0)$ as functions of $1/N_{\max}$. The left panel compares three different fitting functions to extrapolate our results obtained at finite basis sizes to the complete basis by taking the limit $N_{\max} = L_{\max}\to\infty$. The extrapolations using these three functions agree to within $1\%$ of each other. For the remainder of this paper we adopt the second-order polynomial in $1/N_{\max}$ for our extrapolations, fitted to the 4 basis sizes, with an extrapolation uncertainty given by the difference between the result in the largest basis ($N_{\max} = L_{\max}=32$) and the extrapolated value. (Note that this uncertainty does not include any systematic uncertainty coming from the model for the interaction or from the Fock space truncation.) In the right panel of Fig. 4 we show our results for all allowed transitions at finite basis sizes, together with our extrapolation to the complete basis, including our extrapolation uncertainty estimate. Note that they are all close to the nonrelativistic limit 2, which is expected according to our discussion in Sec.~\ref{sec:NR_limit}. For the hindered transitions, the uncertainties from such basis extrapolations are comparatively larger, since their calculations are more sensitive to the details of wavefunctions.
\section{Results}\label{sec:result}
\begin{figure}[htp!]
  \centering
  \includegraphics[width=0.48\textwidth]{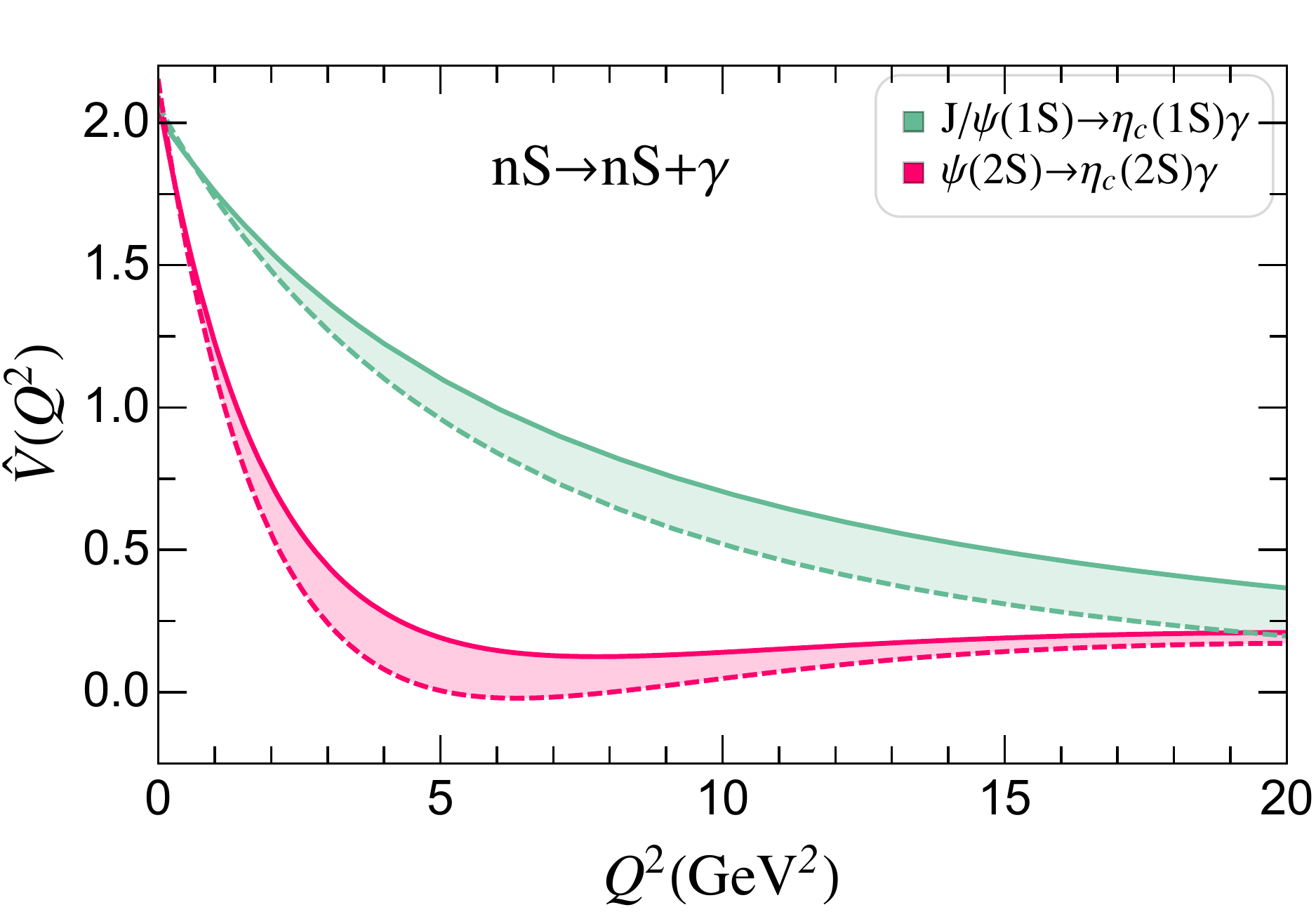}
  \includegraphics[width=0.48\textwidth]{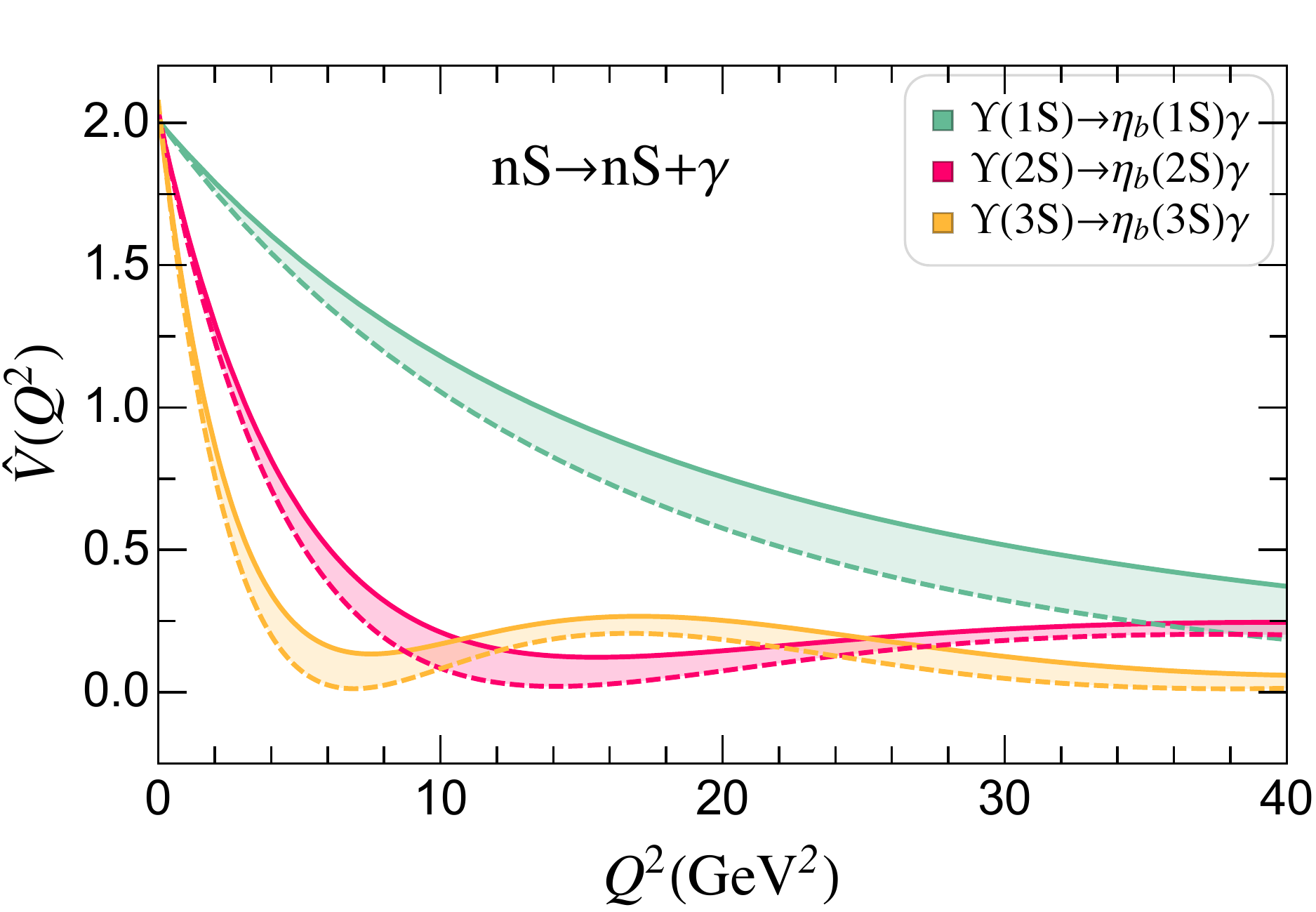}

  \includegraphics[width=0.48\textwidth]{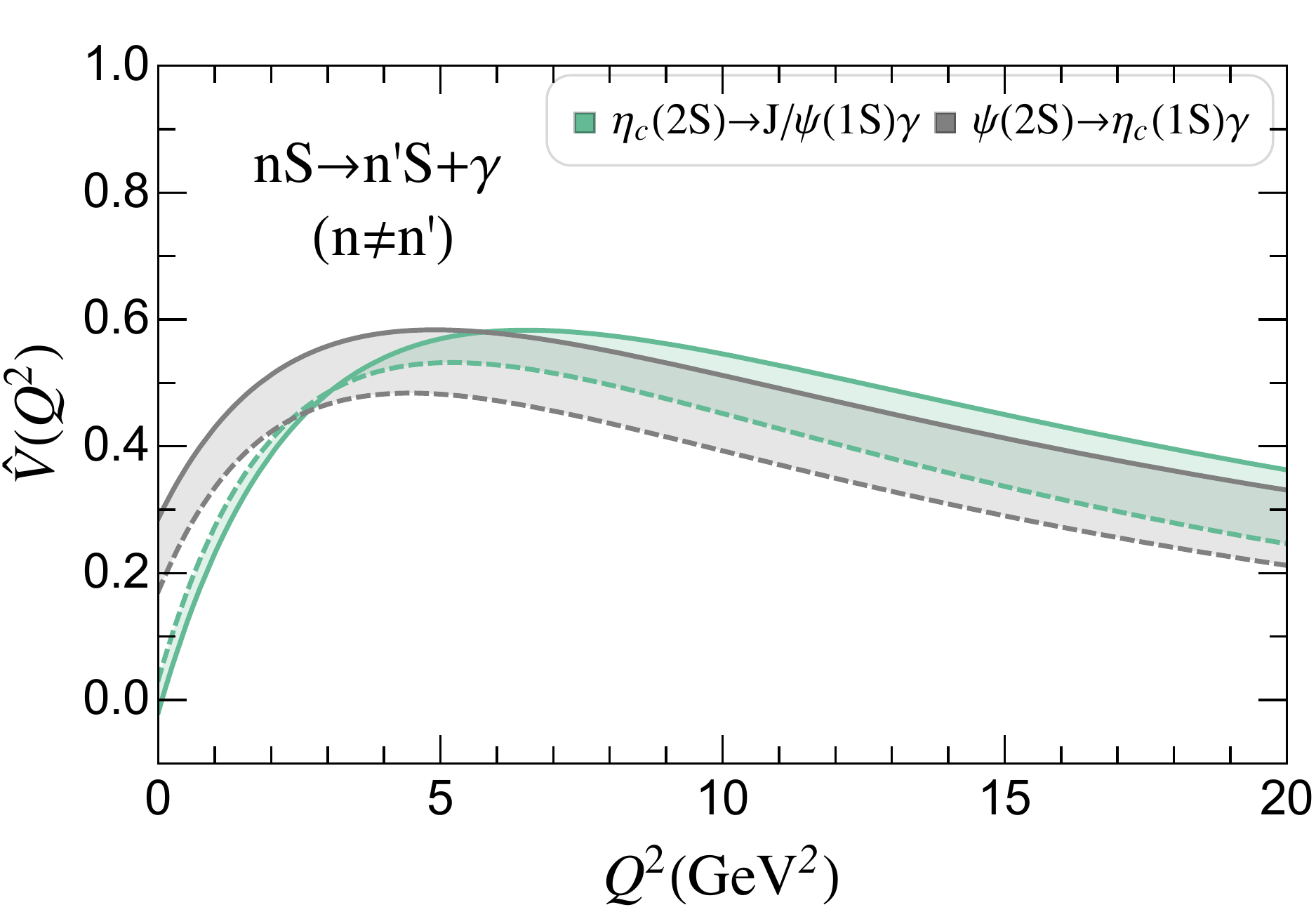}
  \includegraphics[width=0.48\textwidth]{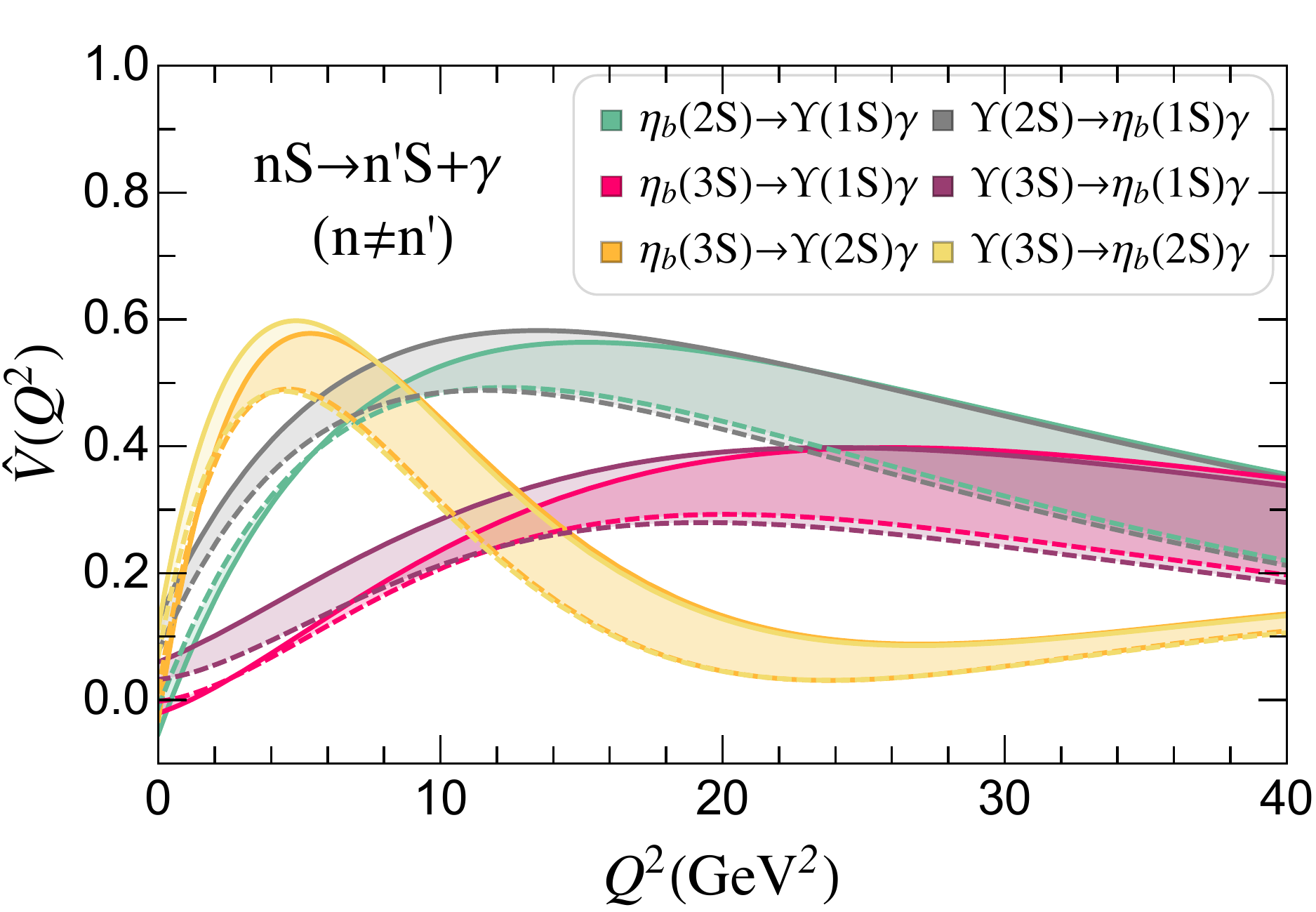}

  \includegraphics[width=0.48\textwidth]{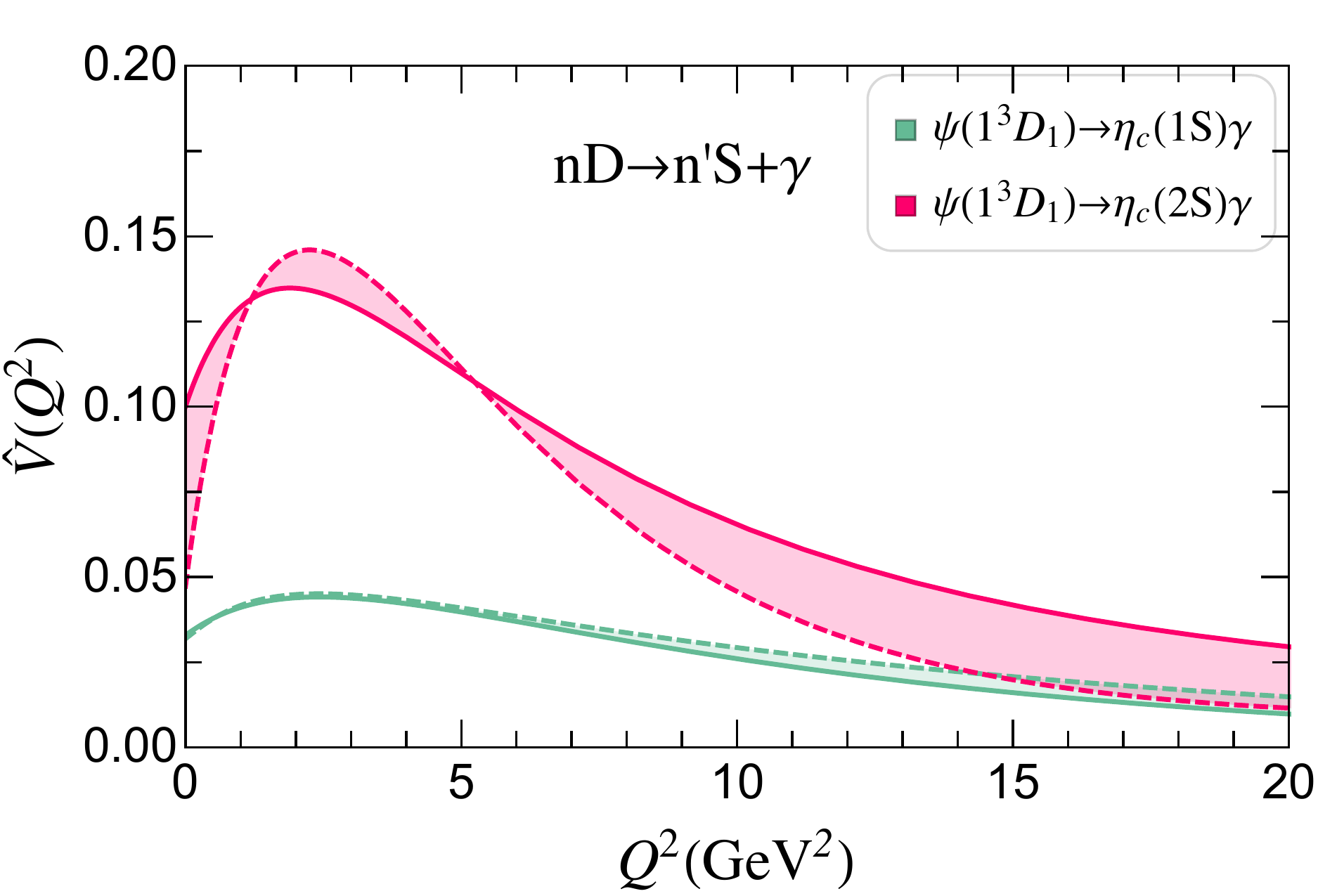}
  \includegraphics[width=0.48\textwidth]{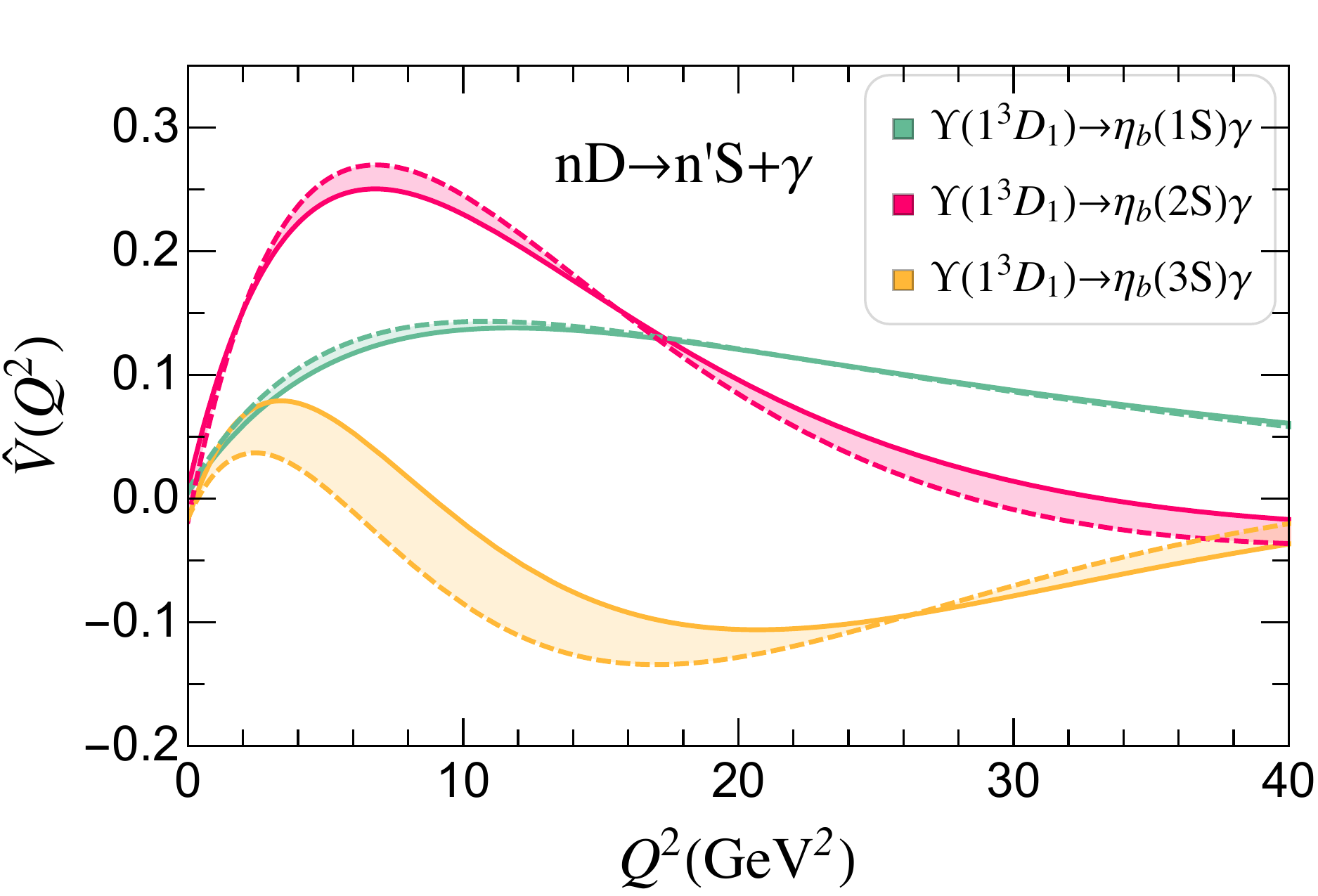}
  \caption{Transition form factors for charmonia (left) and bottomonia (right) are calculated from LFWFs according to Eq.~\eqref{eq:Vmj0}. The first row shows the allowed transitions, the second row shows transitions between different radial excitations, and the third row presents those involving angular excitations. The dashed and solid curves are calculated with LFWFs at $N_{\max}=L_{\max}=8$ and $N_{\max}=L_{\max}=32$ respectively. The shaded areas in between indicates the numerical uncertainty from basis truncation. 
As a consequence of the UV cutoff from the basis, the largest $Q^2 (\simeq\Lambda_{\textsc{uv}}^2)$ at $N_{\max}=32$ truncation 
is $31~{\text{GeV}}^2$ ($44~{\text{GeV}}^2$) for charmonia (bottomonia).}
  \label{fig:TFFcurve}
\end{figure}

Here we present our results for selected pseudoscalar-vector transition form factors for charmonia and bottomonia below their
respective open flavor thresholds. Fig.~\ref{fig:TFFcurve} shows our numerical results of the transition form factors in three groups,
the allowed transition $nS\to nS+\gamma$, the radial excited transition $nS\to n' S+\gamma~ (n\neq n')$ and the angular excited transition $nD\to
n'S+\gamma$, through a progression from upper to lower panels. 
As already discussed in the previous section, for the allowed transitions we find $\hat V(0) \approx 2$, whereas for the hindered transitions involving either radial or angular excitations we have $\hat V(0) \approx 0$.
Transitions involving higher radial excited states feature more wiggles in the curve, which is especially evident in the $nS\to nS+\gamma$ transitions as $n$ increases. This is because the radial excited states have transverse nodes. As a result, the transition form factors, in the form of their convolutions [see Eq.~\eqref{eq:Vmj0}], are not monotonic. The comparison between charmonia and bottomonia is also of interest. For comparable transition modes, the transition form factors show similarity in their patterns as well as their behaviour as a function of $N_{\max}$. 
Furthermore, as illustrated in the second row of panels in Fig.~\ref{fig:TFFcurve}, one observes that the $\mathcal{P}(nS) \to \mathcal{V}(n'S) + \gamma$ transition form factors are very similar to the $\mathcal{V}(nS) \to \mathcal{P}(n'S) + \gamma$ form factors for $n>n'$.

\begin{figure}[htp!]
\begin{flushleft}
  \includegraphics[width=\textwidth]{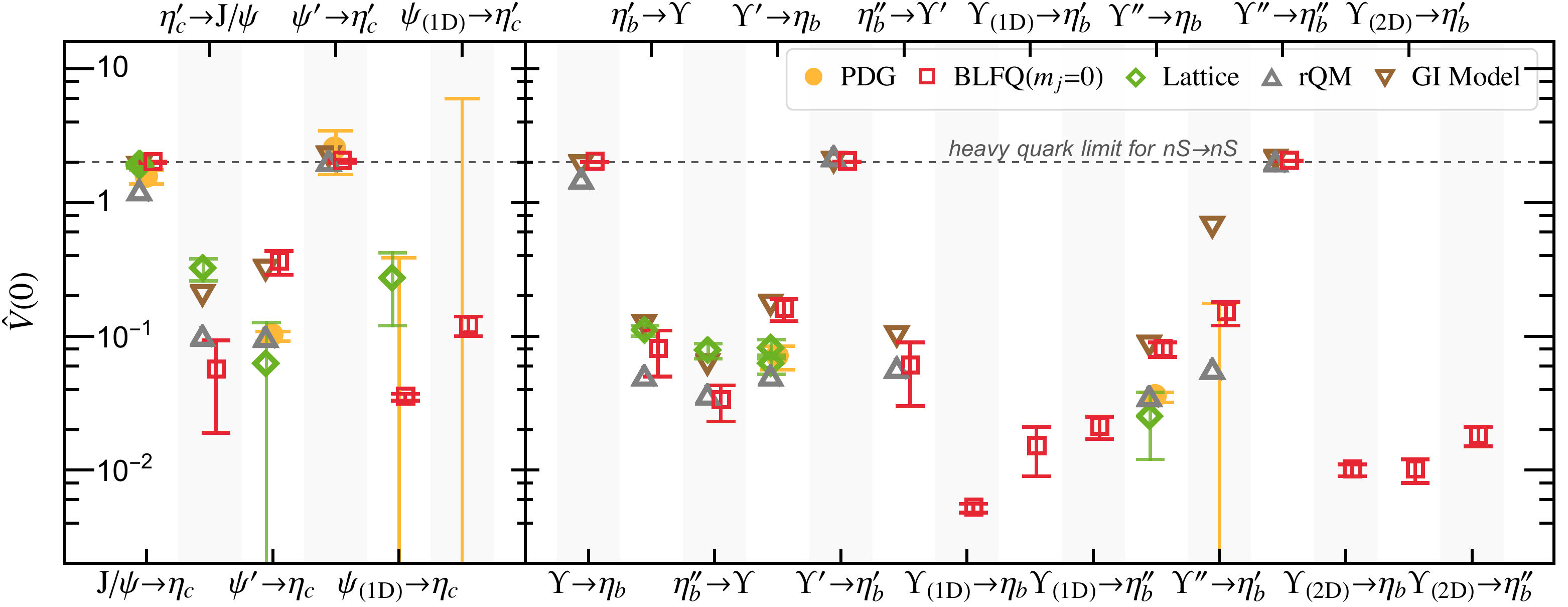}
  \caption{
    $\hat V(0)$ of charmonia and bottomonia transitions, calculated from Eq.~\eqref{eq:Vmj0} and summarized in Table.~\ref{tab:ccbb_Jpl_JR}. Extrapolations are made from
    $N_{\max}=L_{\max}=8,16,24,32$ using second-order polynomials in $N^{-1}_{\max}$. We use the difference between the
    extrapolated and the $N_{\max}=32$ results to quantify numerical uncertainty which is indicated by the vertical error bars on the BLFQ results (sometimes smaller than the symbols). We do not include any systematic uncertainty. Quarkonia in the initial and final states are labeled on the
    top and bottom of the figure. Single (double) apostrophe stands for the radial excited 2S (3S) state. The D-wave states are identified as $n^3D_1$. The heavy quark limit
    $\hat V(0)=2$ of the allowed ($nS\to nS+\gamma$) transition is shown in the dashed line. Results from PDG~\cite{PDG2016},
    Lattice QCD~\cite{Dudek_exotic,Damir2013,Damir2015,Donald_JPsi} and Lattice NRQCD~\cite{PhysRevD.92.094501,Lewis:2012bh}, the relativistic quark model (rQM)~\cite{Bc_NR} and the Godfrey-Isgur (GI) model~\cite{cc_GI,bb_GI} are also presented for
    comparison. 
    }
  \label{fig:ccbbV0}
\end{flushleft}
\end{figure}

Comparisons of $\hat V(0)$ from this work, with experiments (compiled by PDG~\cite{PDG2016}) and with other models (Lattice
QCD~\cite{Dudek_exotic,Damir2013,Damir2015,Donald_JPsi,PhysRevD.92.094501}, Quark Model~\cite{Bc_NR,cc_GI,bb_GI}) are collected in Table~\ref{tab:ccbb_Jpl_JR}
and visualized in Fig.~\ref{fig:ccbbV0}.  
Most calculations, as well as available experimental data, give a value of the the order of 2 for the allowed transitions $nS \to nS + \gamma$: all such data in Table~\ref{tab:ccbb_Jpl_JR} are between 1.5 and 2.5 with only one exception, the relativistic quark model calculation of $J/\psi \to \eta_c + \gamma$. This is in agreement with the vector $\mathcal{V}(nS)$ and the pseudoscalar $\mathcal{P}(nS)$ mesons possessing very similar spatial wavefunctions, but different spin structures.
On the other hand, there is a significant spread in the theoretical results of the hindered transitions.
 This is expected because the hindered transitions involve
changes in radial quantum numbers and/or orbital angular motions and are sensitive to delicate cancellations as discussed above.
Considering the fact that only two free parameters are employed by the model for quarkonia in Ref.~\cite{Yang_run} and the fact that we did not adjust any parameters in our calculation for the transitions,
the agreement to within an order of magnitude is encouraging. 

\begin{figure}[htp!]
  \begin{flushleft}
    \includegraphics[width=\textwidth]{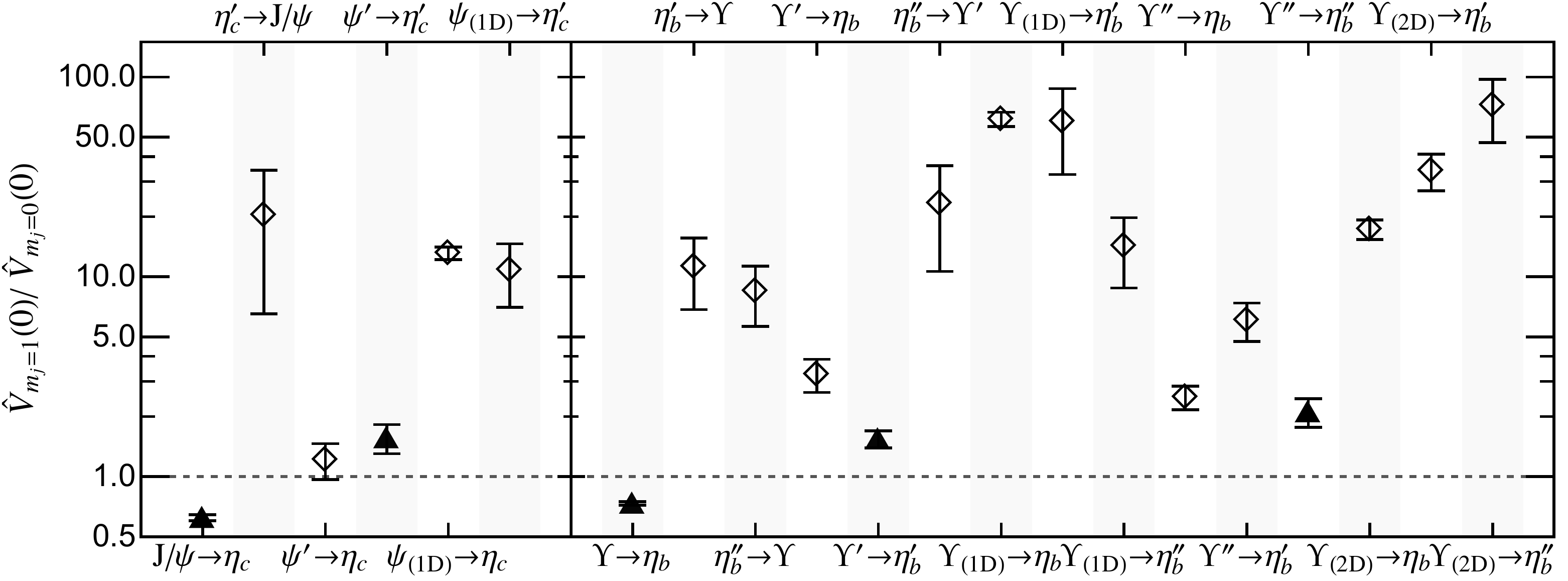}
    \caption{
      Ratio of $\hat V_{m_j=1}(0)$ to $\hat V_{m_j=0}(0)$, calculated from Eq.~\eqref{eq:Vmj1} and Eq.~\eqref{eq:Vmj0} respectively. BLFQ results are
      extrapolated to $N_{\max}=\infty$ from $N_{\max}=L_{\max}=8,16,24,32$ using second-order polynomials in $N^{-1}_{\max}$. We use the difference between the
    extrapolated and the $N_{\max}=32$ results to quantify the numerical uncertainty (indicated by vertical error bars), which is dominated by the uncertainty in the $m_j = 1$ result obtained with the ``+'' component of the current. The $nS\to nS+\gamma$ transitions are shown as filled triangles, whereas the hindered transitions, involving radial/angular excitations, are shown as open diamonds. 
    }
    \label{fig:Vmj_ratio}
  \end{flushleft}
\end{figure}

The results with the ``+'' current, in combination with the $m_j=1$ vector meson wavefunctions, $\hat V_{m_j=1}(0)$, are presented
as a ratio to $\hat V_{m_j=0}(0)$ in Fig.~\ref{fig:Vmj_ratio}.
As already mentioned, we expect these calculations to be much less robust. This is because the calculation of $\hat
V_{m_j=1}(Q^2)$ according to Eq.~\eqref{eq:Vmj1}, depends on subdominant components of the wavefunctions, which is less constrained from the model. Indeed, the dependence of these calculations on the basis truncation is much larger, resulting in significantly larger extrapolation uncertainties. Furthermore, the hindered transitions have
a much larger fluctuation than the allowed transition, due to their sensitivity to the subdominant components in one of the two spatial wavefunctions with different radial quantum numbers and/or different orbital motions.  Our results with the ``+'' component of the current $\hat V_{m_j=1}(0)$, differ by up to 2 orders of magnitude from our more reliable results with the transverse component of the current $\hat V_{m_j=0}(0)$.

\begin{table}[htp!]
\centering
\caption{
      $\hat V(0)$ for radiative decay between $0^{-+}$ and $1^{--}$ charmonia (bottomonia) below the $D\overline{D}$ ($B\overline{B}$) threshold. 
Values from PDG~\cite{PDG2016} are converted from their decay widths according to Eq.~\eqref{eq:VPwidth}. Note that the uncertainties of meson masses propagate into that of $\hat V(0)$.
The BLFQ results are from Eq.~\eqref{eq:Vmj0}. For these results, all meson masses are taken from PDG~\cite{PDG2016}, except that
$\Upsilon(1^3D_1)$, $\Upsilon(2^3D_1)$ and $\eta_b(3S)$ masses are taken from Ref.~\cite{Yang_run}. Extrapolations for BLFQ are
made from $N_{\max}=L_{\max}=8,16,24,32$ using second-order polynomials in $N^{-1}_{\max}$. We use the difference between the
extrapolated and the $N_{\max}=32$ results to quantify numerical uncertainty, which does not include any systematic uncertainty. Uncertainties are quoted in parenthesis and apply to the least significant figures of the result. 
Some lattice results are quoted with more than one source of uncertainty. The lattice nonrelativistic QCD (NRQCD)~\cite{Lewis:2012bh} results are converted from their three-point matrix elements with meson masses from PDG~\cite{PDG2016}.
Values from the relativistic quark model (rQM)~\cite{Bc_NR} and the Godfrey-Isgur (GI) model~\cite{cc_GI,bb_GI} are converted from their decay widths according to Eq.~\eqref{eq:VPwidth} with their suggested meson masses, respectively. 
 These results are plotted in Fig.~\ref{fig:ccbbV0}.
    }
    \vspace{1cm}
  \label{tab:ccbb_Jpl_JR}
    \begin{tabular}{|l|l|l|l|l|l|l|l|l|}
      \hline
      \multirow{2}{*}{$\hat V(0)$}
      & \multirow{2}{*}{PDG~\cite{PDG2016}}
          &BLFQ
      & \multicolumn{4}{l|}{Lattice~\cite{Dudek_exotic,Damir2013,Damir2015,Donald_JPsi,PhysRevD.92.094501,Lewis:2012bh}}
& \multicolumn{2}{l|}{Quark Model}
      \\ \cline{4-9}
      &
          &  ($m_j=0$)
      & Dudek et al.
  & Be{\v{c}}irevi{\'{c}} et al.
      &   HPQCD
& NRQCD
      & rQM~\cite{Bc_NR}
&GI~\cite{cc_GI,bb_GI}
      \\ \hline\hline
      $J/\psi(1S) \to \eta_c(1S)+\gamma$
      &$1.56(19)$
      &$1.99(3)$
      & $1.89(3)$ 
      & $1.92(3)(2)$ 
      &  $1.90(7)(1)$
&
        &1.21
&1.82
      \\ \hline
      $\eta_c(2S) \to J/\psi(1S) +\gamma$
      &
      &$0.056(38)$
      &
      &$0.32(6)(2)$
      &
&
      &0.099
&0.20
      \\\hline
      $\psi(2S)\to \eta_c(1S)+\gamma$
      & $0.100(8) $
      &$0.360(74)$
      & $0.062(64)$
      &
        &
&
      &0.097
&0.31
      \\  \hline
      $\psi(2S)\to \eta_c(2S)+\gamma$
      & $2.52(91)$
      &$ 2.03(6)$
      &
      &
      &
&
      &2.01
&2.18
      \\\hline
      $\psi(1^3D_1)\to\eta_c(1S)  +\gamma$
      & $< 0.377 $
      &$0.035(2) $
      & $0.27(15)$
      &
      &
      &
&
&
      \\\hline
      $\psi(1^3D_1)\to \eta_c(2S) +\gamma$
      & $< 5.84 $
      & $0.121(21)$
      &
      &
      &
      &
&
&
      \\ \hline\hline
      $\Upsilon(1S) \to \eta_b(1S)+\gamma$
      &
      & $2.00(1)$
      &
      &
      &
&
      &1.48
&1.87
    \\ \hline
    $\eta_b(2S) \to \Upsilon(1S) +\gamma$
    &
      & $0.080(27)$
      &
      &
    &
&0.11(1)
      &0.050
&0.12
    \\\hline
    $\eta_b(3S) \to \Upsilon(1S) +\gamma$
    &
      & $0.033(12)$
      &
      &
&
&0.078(10)
      &0.036
&0.061
    \\\hline
    $\Upsilon(2S) \to \eta_b(1S)+\gamma$
    & $0.070(14)$
      & $0.156(30)$
      &
      &
    &  $0.081(13)$
&0.062(10)
      &0.050
&0.17
    \\ \hline
    $\Upsilon(2S) \to \eta_b(2S)+\gamma$
    &
      & $2.01(1)$
      &
      &
    &
&
      &2.17
&1.99
    \\ \hline
    $\eta_b(3S) \to \Upsilon(2S) +\gamma$
    &
      & $0.059(27)$
      &
      &
    &
&
      &0.057
&0.099
    \\\hline
      $\Upsilon(1^3D_1) \to \eta_b(1S)+\gamma$
    &
      & $0.0052(4)$
      &
      &
    &
      &
&
&
    \\ \hline
      $\Upsilon(1^3D_1) \to \eta_b(2S)+\gamma$
    &
      & $0.0148(61)$
      &
      &
    &
      &
&
&
    \\ \hline
    $\Upsilon(1^3D_1) \to \eta_b(3S)+\gamma$
    &
      & $0.021(4)$
      &
      &
    &
      &
&
&
    \\ \hline
    $\Upsilon(3S) \to \eta_b(1S)+\gamma$
    & $0.035(3)$
      & $0.079(10)$
      &
      &
    &
&0.025(13)
      &0.035
&0.084
    \\ \hline
    $\Upsilon(3S) \to \eta_b(2S)+\gamma$
    & $<0.167$
      & $0.145(33)$
      &
      &
    &
&
      &0.056
&0.65
    \\ \hline
    $\Upsilon(3S) \to \eta_b(3S)+\gamma$
    &
      & $2.04(1)$
      &
      &
    &
&
      &1.99
&2.06
    \\ \hline
      $\Upsilon(2^3D_1) \to \eta_b(1S)+\gamma$
    &
      & $0.010(1)$
      &
      &
    &
      &
&
&
    \\ \hline
      $\Upsilon(2^3D_1) \to \eta_b(2S)+\gamma$
    &
      & $0.010(2)$
      &
      &
    &
      &
&
&
    \\ \hline
      $\Upsilon(2^3D_1) \to \eta_b(3S)+\gamma$
    &
      & $0.018(3)$
      &
      &
    &
      &
&
&
    \\ \hline
  \end{tabular}
\end{table}


\section{Decay Constants}\label{sec:cons}
In the discussion of the transition form factors above, we saw that differences could arise when different magnetic projections of
the vector mesons are used, in combination with different components of the electromagnetic current operator. These differences are linked with violations of rotational symmetry in the model.
We argue that the violation of rotational symmetry is not a major factor by checking two representative observables. 
The first one is the meson masses. It has been shown in Ref.~\cite{Yang_run} that the mass eigenvalues associated with different magnetic projections are in
reasonable agreement, with a mean mass spread of 17 MeV (8 MeV) for charmonia (bottomonia). 
The second observable is the decay constant, which we present in this section. The decay constant is defined with the same current operator as the transition form factor. For a vector meson, its value, too, could be extracted from different magnetic projections. On the other hand, it features the simplicity of involving only one LFWF instead of convoluting two LFWFs. Therefore, the decay constant provides a pathway for examining the rotational symmetry of LFWFs.

Decay constants for vector mesons are defined as the local vacuum-to-hadron matrix elements:
\begin{align}
\bra{0} J^\mu (0) \ket{\mathcal{V} (P,m_j)}
    =m_{\mathcal{V}} e^\mu(P, m_j) f_{\mathcal{V}}
    \;.
\end{align}
In Ref.~\cite{Yang_run}, the ``+'' current is used. Since $e^+(P, m_j=\pm 1)=0$, with the ``+'' component of the current the decay constant can only be extracted from $m_j=0$, using
\begin{align}\label{eq:fvmj0}
  f_{\mathcal{V}}(m_j=0)
  =
  \sqrt{2N_c}
  \int_0^1\frac{\diff x}{\sqrt{x(1-x)}}
  \int\frac{\diff^2 \vec k_\perp}{{(2\pi)}^3}
  \psi^{(m_j=0)}_{\uparrow\downarrow+\downarrow\uparrow /\mathcal{V}}(\vec k_\perp, x)
  \;.
\end{align}

However, in analogy to the transition form factor, we can also use the transverse current. In the case of $m_j=0$, we get exactly the
same expression for the decay constant as with 
the ``+'' current, namely Eq.~\eqref{eq:fvmj0}. This should not come as a surprise. Recall that the ``+'' and the transverse matrix elements with the same $m_j$ can be related through a transverse Lorentz boost [see Eq.~\eqref{eq:trLB}]. Furthermore, with the transverse current we can also calculate the decay constant from  the $m_j = \pm 1$ components of the vector meson. The expression for the decay constant with $m_j=1$ follows as 
\begin{align}\label{eq:fvmj1}
  \begin{split}
    f_{\mathcal{V}}(m_j=1)
    =&
    \frac{\sqrt{N_c}}{2m_{\mathcal{V}}}
    \int_0^1\frac{\diff x}{[x(1-x)]^{3/2}}
    \int\frac{\diff^2\vec{k}_\perp}{{(2\pi)}^3}
    [k^L(1-2x)\psi^{(m_j=1)}_{\uparrow\downarrow+\downarrow\uparrow/\mathcal{V}}(\vec k_\perp, x)\\
    &-k^L\psi^{(m_j=1)}_{\downarrow\uparrow-\downarrow\uparrow/\mathcal{V}}(\vec k_\perp, x)
    +\sqrt{2}m_q\psi^{(m_j=1)}_{\uparrow\uparrow/\mathcal{V}}(\vec k_\perp, x)]
    \;.
  \end{split}
\end{align}
Here $m_q$ is the quark mass which is one of the model parameters determined in Ref.~\cite{Yang_run}. Note that using $m_j=-1$ would lead to an equivalent expression considering the symmetry between the $m_j=\pm 1$ LFWFs. 

As is the transition form factor, the decay constant is also Lorentz invariant and thus it should be independent of the
polarization $m_j$. Therefore in practice, the difference between  $f_{\mathcal{V}}(m_j=0)$ and $f_{\mathcal{V}}(m_j=1)$ provides another measure of the violation of rotational symmetry by our model. 
For vector meson states identified as S-wave states, both $f_{\mathcal{V}}(m_j=0)$ and $f_{\mathcal{V}}(m_j=1)$ arise primarily from the dominant spin components of LFWFs, which relate to the nonrelativistic wavefunctions. Moreover, the two expressions reduce to the same form in
the nonrelativistic limit, where $x \to 1/2 + {k_z}/(2m_q)$~\cite{Yang_run} and  $m_q\to m_{\mathcal{V}}/2$. That is, $f_{\mathcal{V}}(m_j=0/1)\to\sqrt{N_c}/(\sqrt{2}m_{\mathcal{V}}) \tilde\psi_{NR}(\vec r=0)$ for S-wave states. 
For vector meson states identified as D-wave states, both $f_{\mathcal{V}}(m_j=0)$ and $f_{\mathcal{V}}(m_j=1)$ are calculated
mainly from the dominant but less occupied spin components of LFWFs (see Sec.~\ref{sec:NR_limit} for discussions of spin
components for D-wave states). The two resulting small values are sensitive to model and numerical uncertainties, and could reveal differences. 
Nevertheless, for the S-wave states of heavy quarkonia, we expect to find robust results when either $m_j=0$
or $m_j=1$ is used to calculate $f_{\mathcal{V}}$. The parameters $m_{\mathcal{V}}$ and $m_q$ involved in $f_{\mathcal{V}}(m_j=1)$ might result in additional uncertainty but the resulting fluctuation should be small for heavy systems.

\begin{figure}[h]
  \centering
  \includegraphics[width=.55\textwidth]{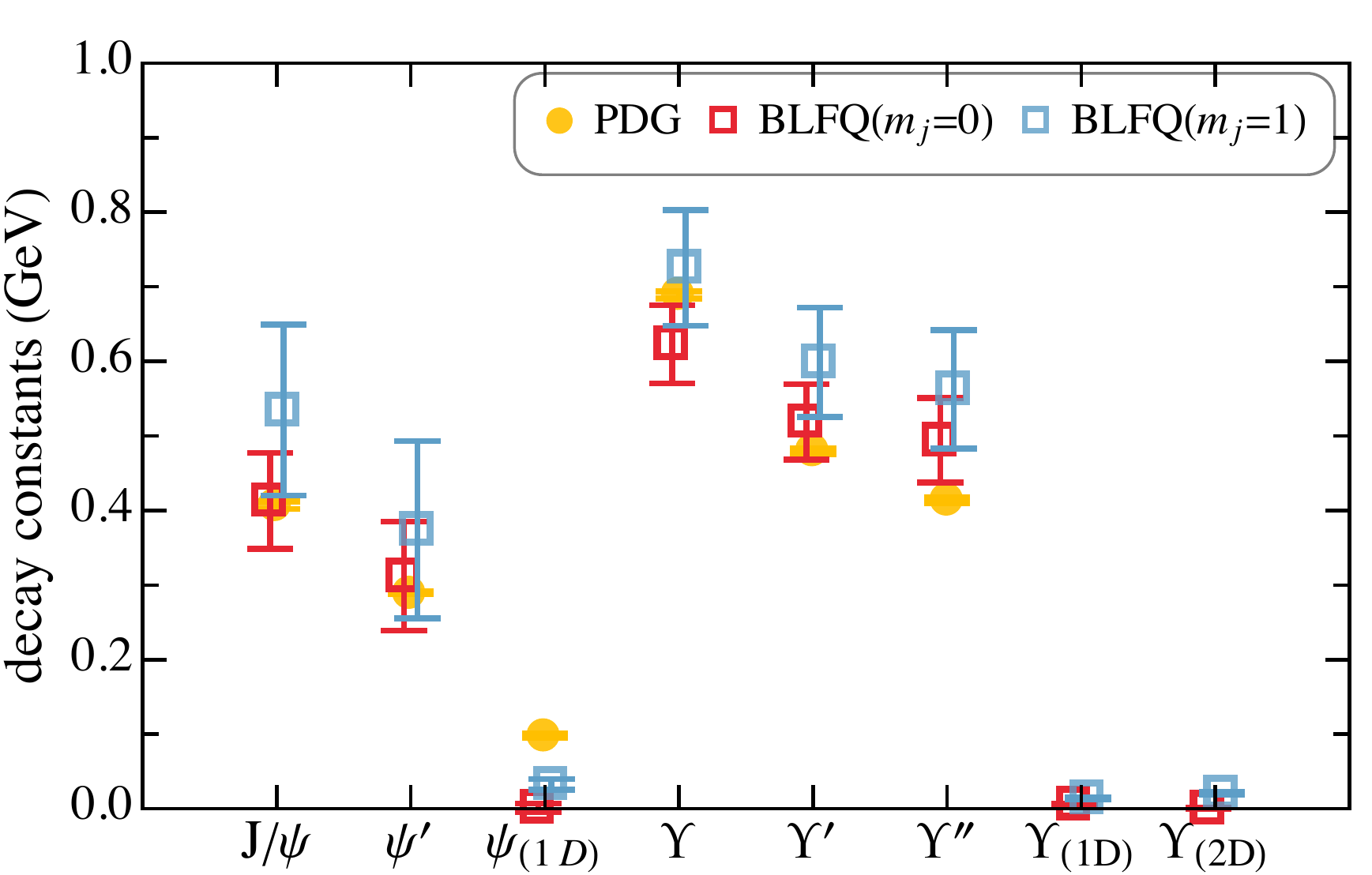}
  \includegraphics[width=.38\textwidth]{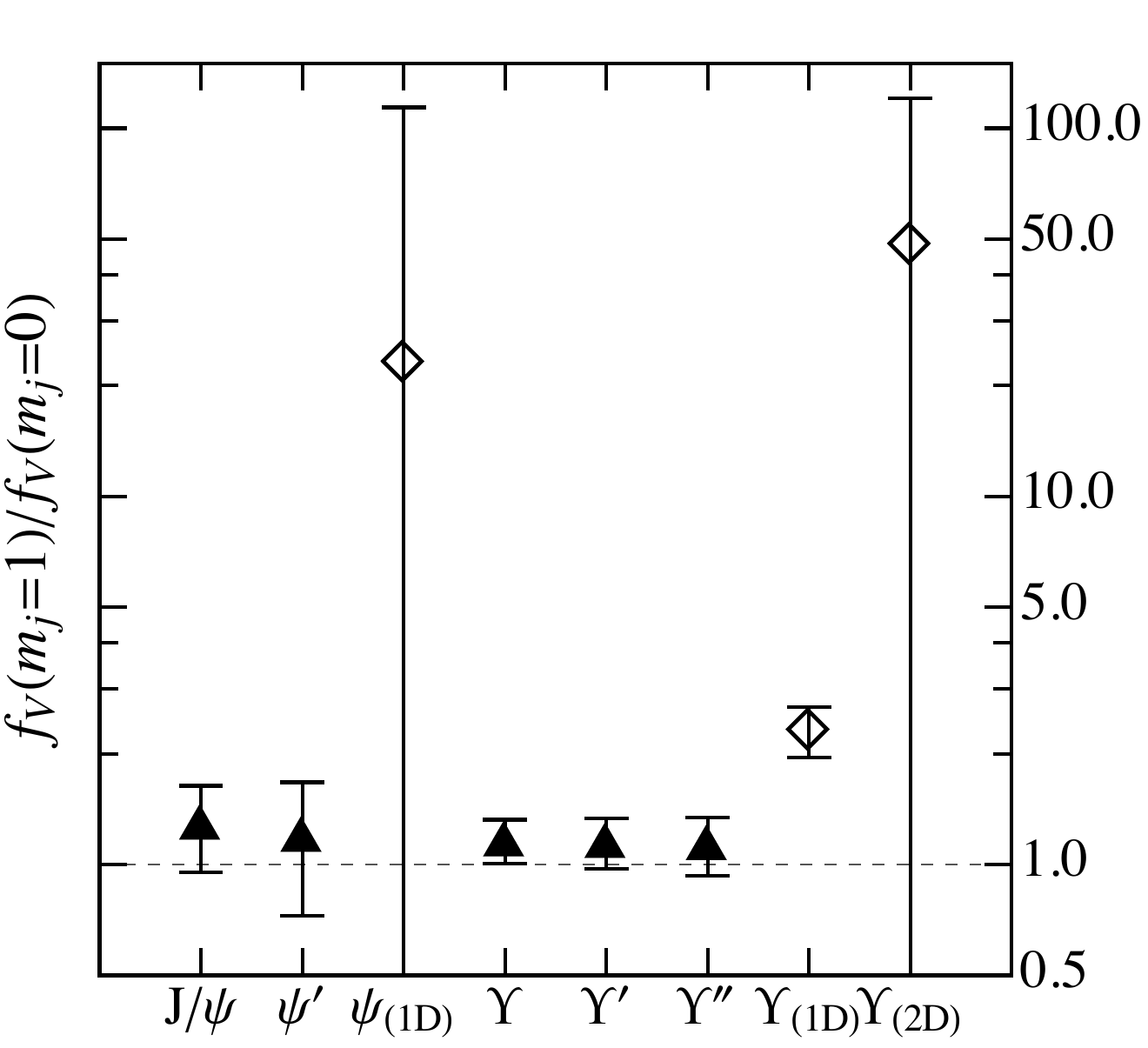}
  \caption{\label{fig:fv} Decay constants of vector heavy quarkonia, calculated from Eqs.~\eqref{eq:fvmj0} and ~\eqref{eq:fvmj1}. Single (double) apostrophe stands for the radial excited 2S (3S) state. The D-wave states are identified as $n^3D_1$. The results are obtained with $N_{\max}=L_{\max}=8$ with error bars $\Delta
    f_{c\bar{c}}=|f_{c\bar{c}}(N_{\max}=8)-f_{c\bar{c}}(N_{\max}=16)|$ for charmonium, and $N_{\max}=L_{\max}=32$ with error bars $\Delta
    f_{b\bar{b}}=2|f_{b\bar{b}}(N_{\max}=32)-f_{b\bar{b}}(N_{\max}=24)|$ for bottomonium.  
Results from PDG~\cite{PDG2016} are provided
    for comparison. The right panel shows the ratio of $f_{\mathcal{V}}(m_j=1)$ to $f_{\mathcal{V}}(m_j=0)$, where the S-wave states are shown in filled triangles and the D-wave states are shown in open diamonds.}
\end{figure}

We use both Eqs.~\eqref{eq:fvmj0} and ~\eqref{eq:fvmj1} to calculate the decay constants for the lowest three charmonium and five bottomonium vector states below their open flavor thresholds. The results are presented in Fig.~\ref{fig:fv}, where basis truncations are chosen to match the UV cutoffs $\Lambda_{\textsc{uv}}\simeq  \kappa \sqrt{N_{\max}} \approx 1.7 m_q$~\cite{Yang_run}. The two sets of results,
$f_{\mathcal{V}}(m_j=0)$ and $f_{\mathcal{V}}(m_j=1)$, are all within each others' extrapolation uncertainties for the five S-wave states in Fig.~\ref{fig:fv}. Such consistency implies that the
rotational symmetry is reasonably preserved in LFWFs. For the D-wave states, the decay constants are small but differences between $f_{\mathcal{V}}(m_j=0)$ and $f_{\mathcal{V}}(m_j=1)$ can be noticed since each result depends on different small components.
By implication, the transition form factor $\hat V(Q^2)_{m_j=0}$
calculated using the transverse current, with its overlapping dominant components of both the initial and the final LFWFs, is further supported as a robust result. 

\section{Summary and Discussions}\label{sec:summary}
We have derived formulae for the radiative transitions between vector meson and pseudoscalar mesons (M1 transitions) in light-front
dynamics. We then calculated the transition form factors for heavy quarkonia obtained in the BLFQ approach. The
majority of the predictions from the BLFQ approach are in reasonable agreement with experimental data and other model calculations when the
$m_j=0$ state of the vector meson and the transverse current $J^R=J^x+i J^y$ are employed. We have also shown that
at least in the context of heavy quarkonium, this choice is preferred to the traditional good current $J^+$ (cf.~Ref.~\cite{CARBONELL1998215}).
The latter lacks the access to the $m_j=0$ state of the vector mesons, and generates a less robust result.
Moreover, in extracting the decay constants of vector mesons, $J^+$
gives the same result as $J^R$ when the $m_j=0$ state is used, but can not be used when the $m_j=1$ state is taken.

The comparison of results extracted from different $m_j$ states of the vector mesons also allowed us to investigate the rotational symmetry of the
model. The consistency of decay constants, $f_{\mathcal{V}}(m_j=0)$ and $f_{\mathcal{V}}(m_j=1)$, suggests that the rotational symmetry is achieved
at a reasonable level with the adopted LFWFs for states identified as S-waves. 
By contrast, the M1 transition form factors, $\hat V(Q^2)_{m_j=0}$ and $\hat V(Q^2)_{m_j=1}$, as convolutions of  the initial and final
states LFWFs, reveal details of the violation of symmetry between $m_j=0$ and $m_j=1$ LFWFs, even between S-wave states. In such circumstance, $\hat V(Q^2)_{m_j=0}$ is found to be more robust due to
overlapping dominant parts of LFWFs, and it became our suggested approach in heavy systems.
However, for states identified as D-waves, our results for both the transition form factors and for the decay constants show large
difference between calculations using $m_j=0$ and calculations with $m_j=1$. This suggests that our $m_j=0$ and $m_j=1$ LFWF for these states
have noticeably different contributions beyond the D-waves, such as those from the S-waves (which can be present even in the
nonrelativistic calculations) and those with purely relativistic origin. In addition, they could also have different components
that vanish when rotational symmetry is restored.
The breaking of rotational symmetry in the wavefunction mainly results from the Fock sector truncation and basis
truncation. We expect to improve rotational symmetry by including higher Fock sectors, as shown in some simpler theories~\cite{Karmanov:2012aj_FDRN}.

In future work, we will also extend the current formalism to electric dipole transitions (E1 transitions) and higher multipole
contributions such as M2. With a variety of transition modes, we hope to establish a more comprehensive understanding on the internal structure of
quark-antiquark bound states and to further address the role of Lorentz symmetry. Radiative transitions within other meson sectors, such as
the light meson system and the mixed flavor systems, are also of great interest. Furthermore, besides the currently used Drell-Yan
frame, we could also adopt various other frames where the transition form factor in the time-like region may be accessible. The frame dependence study could also serve as a metric for the violation of the Lorentz symmetry, as shown for the elastic form factors~\cite{Li:2017uug}.
\section*{Acknowledgements}
We wish to thank Lekha Adhikari, Guangyao Chen, Shaoyang Jia, V.A. Karmanov, Sofia Leit{\~a}o, Wayne N. Polyzou, Wenyang Qian,
Shuo Tang,  Anji Yu and Xingbo Zhao for valuable discussions. We thank R.M.Woloshyn for his communications regarding his calculation on the three-point matrix elements.
This work was supported in part by the US Department of Energy (DOE) under Grant Nos. DE-FG02-87ER40371, DE-SC0018223
(SciDAC-4/NUCLEI), DE-SC0015376 (DOE Topical Collaboration in Nuclear Theory for Double-Beta Decay and Fundamental
Symmetries) and DE-FG02-04ER41302. A portion of the computational resources were provided by the National Energy Research Scientific Computing Center
(NERSC), which is supported by the US DOE Office of Science.

\appendix

\section{Conventions}\label{sec:convention}
We adopt the conventions of Ref.~\cite{Yang_run}. Here we provide some additional identities useful for this work. 
\subsection{Light-Front coordinates}
The metric
\begin{align}
  g^{+-}=g^{-+}=2, \qquad 
g_{+-}=g_{-+}=\frac{1}{2}, 
\qquad g^{xx}=g^{yy}=-1 \;.
\end{align}
So $\sqrt{-\det g}=\frac{1}{2}$.
The Levi-Civita tensor should be defined as 
\begin{align}
  \epsilon^{\mu\nu\rho\sigma}=\frac{1}{\sqrt{-\det g}}
\begin{cases}
+1, &\text{if $\mu,\nu,\rho,\sigma$ is an even permutation of $-,+,x,y$}\\
-1,&\text{if $\mu,\nu,\rho,\sigma$ is an odd permutation of $-,+,x,y$}\\
0,&\text{other cases}
\end{cases}
    \;.
\end{align}

\subsection{Spinor matrix elements}\label{sec:spinors}

\begin{align}\label{eq:ubgmu}
\begin{split}
  &\bar{u}_{s'}(p_2)\gamma^+ u_s(p_1)=2\sqrt{p_1^+ p_2^+}\delta_{s',s}\\
  &\bar{u}_{s'}(p_2)\gamma^- u_s(p_1)
=\frac{2}{\sqrt{p_1^+ p_2^+}}
\begin{cases}
  m_q^2+p_1^R p_2^L, &s=+, s'=+\\
  m_q(p_2^L -p_1^L), &s=-,  s'=+\\
  m_q(p_1^R -p_2^R), &s=+,  s'=-\\
  m_q^2+p_1^L p_2^R, &s=-, s'=-
\end{cases}\\
&\bar{u}_{s'}(p_2)\gamma^R u_s(p_1)
=\frac{2}{\sqrt{p_1^+ p_2^+}}
\begin{cases}
  p_2^+p_1^R, &s=+, s'=+\\
  m_q(p_2^+ -p_1^+), &s=-,  s'=+\\
  0, &s=+,  s'=-\\
  p_1^+ p_2^R, &s=-, s'=-
\end{cases}\\
&\bar{u}_{s'}(p_2)\gamma^L u_s(p_1)
=\frac{2}{\sqrt{p_1^+ p_2^+}}
\begin{cases}
p_1^+ p_2^L, &s=+, s'=+\\
 0, &s=-,  s'=+\\
  m_q(p_1^+ -p_2^+), &s=+,  s'=-\\
  p_2^+p_1^L, &s=-, s'=-
\end{cases}
\end{split}
\end{align}

\begin{align}
  \bar{v}_s(p_1)\gamma^\mu v_{s'}(p_2)=\bar{u}_{s'}(p_2)\gamma^\mu u_s(p_1)
  \end{align}

\bibliography{RdDcy}

\begin{thebibliography}{44}
\bibitem [{\citenamefont {Tang}\ and\ \citenamefont {et~al.}()}]{Shuo_Bc}%
  \BibitemOpen
  \bibfield  {author} {\bibinfo {author} {\bibfnamefont {S.}~\bibnamefont
  {Tang}}\ and\ \bibinfo {author} {\bibnamefont {et~al.}},\ }\href@noop {}
  {\enquote {\bibinfo {title} {${B}_{c}$ and its properties on the light
  front},}\ }\bibinfo {note} {In preparation}\BibitemShut {NoStop}%
\end{thebibliography}
\end{document}